\documentclass{sig-alternate}
\usepackage{graphicx}
\usepackage[caption=false,hangindent=6pt,margin=2pt]{subfig}
\usepackage[hidelinks]{hyperref}
\usepackage{color,comment,xparse}
\usepackage{tikz}
\usetikzlibrary{shapes.arrows,shapes.geometric,positioning,fit,calc,patterns}

\graphicspath{{./}{eps/}{png/}}

\newcommand{\indr}[1]{\mathbf{1}\left( #1 \right)}
\NewDocumentCommand{\E}{o m}{\mathbb{E}\IfValueTF{#1}{_{#1}}{}\left[ #2 \right]}

\newcommand{\Guu}{\mathcal{G}_\text{uu}}
\newcommand{\Guc}{\mathcal{G}_\text{uc}}
\newcommand{\Euu}{\mathcal{E}_\text{uu}}
\newcommand{\Euc}{\mathcal{E}_\text{uc}}
\newcommand{\CL}{\mathcal{L}}

\newtheorem{theorem}{Theorem}

\newcommand{\header}[1]{\smallskip\noindent\textbf{#1}}

\begin{document}

\title{Tracking Triadic Cardinality Distributions for Burst Detection in
Social Activity Streams}

\numberofauthors{1}

\author{
\alignauthor
Junzhou Zhao\textsuperscript{*}\quad
John C.S. Lui\textsuperscript{\dag}\quad
Don Towsley\textsuperscript{\ddag}\quad
Pinghui Wang\textsuperscript{\S}\quad
Xiaohong Guan\textsuperscript{*}\\
\smallskip
\affaddr{\textsuperscript{*} Xi'an Jiaotong University, China\\
\textsuperscript{\dag} The Chinese University of Hong Kong, Hong Kong\\
\textsuperscript{\ddag} University of Massachusetts at Amherst, US\\
\textsuperscript{\S} Huawei Noah's Ark Lab, Hong Kong}\\
\email{\{jzzhao, xhguan\}@sei.xjtu.edu.cn,
cslui@cse.cuhk.edu.hk,
towsley@cs.umass.edu,
wang.pinghui@huawei.com}
}

\maketitle

\begin{abstract}
In everyday life, we often observe unusually frequent interactions among
people before or during important events, e.g., we receive/send more
greetings from/to our friends on Christmas Day, than usual.
We also observe that some videos suddenly go viral through people's sharing
in online social networks (OSNs).
Do these seemingly different phenomena share a common structure?

All these phenomena are associated with sudden surges of user activities
in networks, which we call ``\emph{bursts}'' in this work.
We find that the emergence of a burst is accompanied with the formation of
triangles in networks.
This finding motivates us to propose a new method to detect bursts in OSNs.
We first introduce a new measure, ``{\em triadic cardinality
  distribution}'', corresponding to the fractions of nodes with different
numbers of triangles, i.e., triadic cardinalities, within a network.
We demonstrate that this distribution changes when a burst occurs, and is
naturally immunized against spamming social-bot attacks.
Hence, by tracking triadic cardinality distributions, we can reliably
detect bursts in OSNs.
To avoid handling massive activity data generated by OSN users, we design
an efficient sample-estimate solution to estimate the triadic cardinality
distribution from sampled data.
Extensive experiments conducted on real data demonstrate the usefulness of
this triadic cardinality distribution and the effectiveness of our
sample-estimate solution.

\end{abstract}

\header{Categories and Subject Descriptors:}
J.4 {\bf[Computer Applications]}: Social and Behavioral Sciences

\header{General Terms:}
Design, Measurement

\header{Keywords:}
Social Activity Streams, Burst Detection, Sampling Methods,
Data Stream Algorithms

\section{Introduction}
\label{sec:intro}

Online social networks (OSNs) have become ubiquitous platforms that provide
various ways for users to interact over the Internet, such as tweeting
tweets, sharing links, messaging friends, commenting on posts, and
mentioning/replying to other users (i.e., @someone).
When intense user interactions take place in a short time period,
there will be a surge in the volume of user activities in an OSN.
Such a surge of user activity, which we call a \emph{burst} in this work,
usually relates to emergent events that are occurring or about to occur in
the real world.
For example, Michael Jackson's death on June 25, 2009 triggered a global
outpouring of grief on Twitter~\cite{Harvey2009}, and the event even
crashed Twitter for several minutes~\cite{Shiels2009}.
In addition to bursts caused by real world events, some bursts arising from
OSNs can also cause enormous social impact in the real world.
For example, the 2011 England riots, in which people used OSNs to organize,
resulted in $3,443$ crimes across London due to this
disorder~\cite{LondonRiots2011}.
Hence, detecting bursts in OSNs is an important task,
both for OSN managers to monitor the operation status of an OSN,
and for government agencies to anticipate any emergent social disorder.

Typically, there are two types of user interactions in OSNs.
First is the interaction between users (we refer to this as {\em user-user
  interaction}), e.g., a user sends a message to another user, while the
second is the interaction between a user and a media content piece (we
refer to this as {\em user-content interaction}), e.g., a user posts a
video link.
Examples of bursts caused by these two types of interactions include,
many greetings being sent/received among people on Christmas Day,
and videos suddenly becoming viral after one day of sharing in an OSN.
At first sight, detecting such bursts in an OSN is not difficult.
For example, a straightforward way to detect bursts caused by user-user
interactions is to {\em count} the number of pairwise user interactions
within a time window, and report a burst if the volume lies above a given
threshold.
However, this method is vulnerable to spamming social-bot attacks~\cite{Chu2010,Grier2010,Stringhini2010a,Boshmaf2011,Thomas2011,Beutel2013a}, which
can suddenly generate a huge amount of spamming interactions in the OSN.
Hence, this method can result in many {\em false alarms} due to the
existence of social bots.
Similar problem also exist when detecting bursts caused by user-content
interactions.
Many previous works on burst detection are based on idealistic
assumptions~\cite{Kleinberg2002,Yi2005,Parikh2008,Eftekhar2013} and
simply ignore the existence of social bots.

\header{Present work.}
The primary goal of this work is to leverage a special {\em triangle
  structure}, which is a feature of humans, to design a robust burst
detection method that is immunized against common social-bot attacks.
We first describe the triangle structure shared by both types of user
interactions.

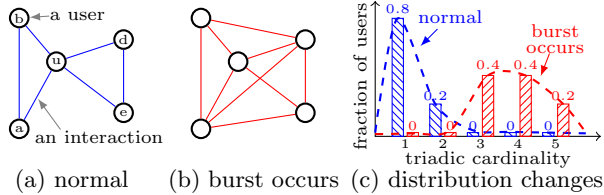
\begin{figure}
  \centering
\tikzstyle{usr}=[draw,thick,circle,minimum size=7pt,inner sep=0]
\tikzstyle{bbar}=[blue,pattern=north west lines,pattern color=blue]
\tikzstyle{rbar}=[red,pattern=north east lines,pattern color=red]
\tikzstyle{line}=[dashed,thick]
\tikzstyle{txt}=[inner sep=0,anchor=west,align=center]
\tikzstyle{arr}=[->,>=latex,gray]

\subfloat[normal\label{f:ib1}]{
\begin{tikzpicture}
\tiny
\path[use as bounding box] (0.2,-.1) rectangle (2.4,2);

\node(a) [usr] at (.6,.3) {a};
\node(b) [usr, above = 1.2cm of a] {b};
\node(c) [usr, above right = .7cm and .3cm of a] {u};
\node(d) [usr, above right = .1cm and .7cm of c] {d};
\node(e) [usr, below = .7cm of d] {e};
\draw[blue] (a) -- (b) -- (c) -- (d) -- (e) -- (c)-- (a);

\node(au) [txt] at (1,1.8) {\scriptsize a user};
\node(ai) [txt] at (.8,.2) {\scriptsize an interaction};
\draw[arr] (au.west) -- (b);
\draw[arr] (ai.170) -- (.85,.7);

\end{tikzpicture}
}%
\subfloat[burst occurs\label{f:ib2}]{
\begin{tikzpicture}
\tiny
\path[use as bounding box] (0.2,-.1) rectangle (2.4,2);

\node(a) [usr] at (.6,.3) {};
\node(b) [usr, above = 1.2cm of a] {};
\node(c) [usr, above right = .7cm and .3cm of a] {};
\node(d) [usr, above right = .1cm and .7cm of c] {};
\node(e) [usr, below = .7cm of d] {};
\draw[red] (a) -- (b) -- (c) -- (a) -- (e) -- (c) -- (d) -- (e);
\draw[red] (b) -- (d) -- (a);
\end{tikzpicture}
}%
\subfloat[distribution changes\label{f:ib3}]{
\begin{tikzpicture}
\tiny
\path[use as bounding box] (-.3,-.2) rectangle (3,1.9);

\draw[<->] (0,1.9) -- (0,0.1) -- (3,0.1);
\foreach \x in {1,2,3,4,5} \node (\x) at (\x*0.5-0.1,0) {$\x$};
\node(pdf) [rotate=90] at (-.2,.9) {\scriptsize fraction of users};
\node[txt,anchor=center] at (1.5,-0.2) {\scriptsize triadic cardinality};

\begin{scope}[yshift=1mm]
\draw[bbar] (0.225,0) rectangle (0.375,1.57); 
\draw[bbar] (0.725,0) rectangle (0.875,0.43); 
\draw[bbar] (1.225,0) rectangle (1.375,0.05); 
\draw[bbar] (1.725,0) rectangle (1.875,0.05); 
\draw[bbar] (2.225,0) rectangle (2.375,0.05); 
\draw[line,blue] plot[smooth] coordinates {(0,0.05) (0.3,1.57)
(0.8,0.43) (1.3,0.08) (1.8,0.06) (2.3,0.05) (2.8,0.04)};
\node[blue] at (1,1.6) (ex1) {\scriptsize normal};
\draw[arr,blue] (ex1) -- (0.6,1.2);
\node[blue] at (0.3,1.7) {$0.8$};
\node[blue] at (0.8,0.54) {$0.2$};
\node[blue] at (1.3,0.145) {$0$};
\node[blue] at (1.8,0.145) {$0$};
\node[blue] at (2.3,0.145) {$0$};
\draw[rbar]  (0.425,0) rectangle (0.575,0.05); 
\draw[rbar]  (0.925,0) rectangle (1.075,0.05); 
\draw[rbar]  (1.425,0) rectangle (1.575,0.81); 
\draw[rbar]  (1.925,0) rectangle (2.075,0.81); 
\draw[rbar]  (2.425,0) rectangle (2.575,0.43); 
\draw[line,red] plot[smooth] coordinates { (0,0.03) (0.5,0.04)
(1.0,0.08) (1.5,0.81) (2.0,0.81) (2.5,0.43) (2.8,0.05)};
\node[red,txt] at (2,1.3) (ex) {{\scriptsize burst} \\ {\scriptsize occurs}};
\draw[arr,red] (ex) -- (2.2,0.8);
\node[red] at (0.5,0.145) {$0$};
\node[red] at (1.0,0.145) {$0$};
\node[red] at (1.5,0.95) {$0.4$};
\node[red] at (2.0,0.95) {$0.4$};
\node[red] at (2.5,0.54) {$0.2$};
\end{scope}
\end{tikzpicture}
}

  \vspace{-5pt}
  \caption{Interaction burst and interaction triangle}
  \label{fig:ib_exam}
  \vspace{-8pt}
\end{figure}

\header{Interaction triangles in user-user interactions.}
Humans form social networks with larger clustering coefficients
than those in random networks~\cite{Watts1998} because social
networks exhibit many \emph{triadic closures}~\cite{Kossinets2006}.
This is due to the social phenomenon of ``{\em friends of my friends are
  also my friends}''.
Since user-user interactions usually take place along social links, this
property implies that user-user interactions should also exhibit many
triadic closures (which we will verify in experiments).
In other words, when a group of users suddenly become active, or we say an
{\em interaction burst} occurs, in addition to observing the rise of volume of
pairwise interactions, we expect to also observe many interactions among three
neighboring users, i.e., many {\em interaction triangles} form if we
consider an edge of an interaction triangle to be a user-user interaction.
This is illustrated in Fig.~\ref{f:ib1} when no interaction burst occurs,
while in Fig.~\ref{f:ib2}, an interaction burst occurs.
In contrast, activities generated by social bots do not possess many
triangles since social bots typically select their targets randomly from an
OSN~\cite{Boshmaf2011,Thomas2011}.

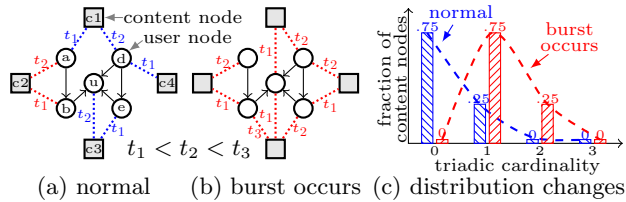
\begin{figure}
  \centering
\tikzstyle{usr}=[draw,thick,circle,minimum size=7pt,inner sep=0]
\tikzstyle{csd}=[draw,thick,rectangle,fill=gray!20,minimum size=7pt,inner sep=0]
\tikzstyle{csdlne}=[densely dotted,thick]
\tikzstyle{lb}=[midway,inner sep=#1]
\tikzset{lb/.default=1pt}
\tikzstyle{bbar}=[blue,pattern=north west lines,pattern color=blue]
\tikzstyle{rbar}=[red,pattern=north east lines,pattern color=red]
\tikzstyle{curve}=[dashed,thick]
\tikzstyle{txt}=[inner sep=0,anchor=west,text depth=\depthof{g},
align=center]
\tikzstyle{arr}=[gray,->,>=latex]

\newcommand{\tm}[1]{$t_{\textsc{#1}}$}

\subfloat[normal\label{f:cb1}]{
\begin{tikzpicture}
\tiny
\path[use as bounding box] (0.15,-.1) rectangle (2.35,2);
\node(f) [usr] at (1.25,1) {u};
\node(a) [usr, above left = 4pt and 5pt of f] {a};
\node(b) [usr, below left = 4pt and 5pt of f] {b};
\node(d) [usr, above right = 4pt and 5pt of f] {d};
\node(e) [usr, below right = 4pt and 5pt of f] {e};
\draw[->] (a) -- (b);
\draw[->] (b) -- (f);
\draw[->] (d) -- (e);
\draw[->] (e) -- (f);
\draw[->] (d) -- (f);
\node(c1) [csd, above = 6mm of f] {c1};
\node(c2) [csd, left = 7mm of f] {c2};
\node(c3) [csd, below = 6mm of f] {c3};
\node(c4) [csd, right = 7mm of f] {c4};
\node[txt,right=3mm of c3] {\footnotesize $t_1<t_2<t_3$};

\draw[blue,csdlne](a)--(c1) node[lb=0,left,pos=.6]{\tm{1}};
\draw[blue,csdlne](d)--(c1) node[lb,right,pos=.6]{\tm{2}};
\draw[red,csdlne](a)--(c2) node[lb,above,pos=.6]{\tm{2}};
\draw[red,csdlne](b)--(c2) node[lb,below,pos=.6]{\tm{1}};
\draw[blue,csdlne](f)--(c3) node[lb=0,left]{\tm{2}};
\draw[blue,csdlne](e)--(c3) node[lb,right,pos=.6]{\tm{1}};
\draw[blue,csdlne](d)--(c4) node[lb,above,pos=.7]{\tm{1}};

\node(ac) [txt,right=.25 of c1] {\scriptsize content node};
\node(au) [txt] at (1.9,1.6) {\scriptsize user node};
\draw[arr] (ac.west) -- (c1.east);
\draw[arr] (au.183) -- (d.north east);

\end{tikzpicture}
}%
\subfloat[burst occurs\label{f:cb2}]{
\begin{tikzpicture}
\tiny
\path[use as bounding box] (0.15,-.1) rectangle (2.35,2);
\node(f) [usr] at (1.25,1) {};
\node(a) [usr, above left = 4pt and 5pt of f] {};
\node(b) [usr, below left = 4pt and 5pt of f] {};
\node(d) [usr, above right = 4pt and 5pt of f] {};
\node(e) [usr, below right = 4pt and 5pt of f] {};
\draw[->] (a) -- (b);
\draw[->] (b) -- (f);
\draw[->] (d) -- (e);
\draw[->] (e) -- (f);
\draw[->] (d) -- (f);
\node(c1) [csd, above = 6mm of f] {};
\node(c2) [csd, left = 7mm of f] {};
\node(c3) [csd, below = 6mm of f] {};
\node(c4) [csd, right = 7mm of f] {};

\draw[red,csdlne](f)--(c1) node[lb=0,left]{\tm{1}};
\draw[red,csdlne](d)--(c1) node[lb,right]{\tm{2}};
\draw[red,csdlne](a)--(c2) node[lb,above,pos=.6]{\tm{2}};
\draw[red,csdlne](b)--(c2) node[lb,below,pos=.6]{\tm{1}};
\draw[red,csdlne](b)--(c3) node[lb=0,left,pos=.7]{\tm{3}};
\draw[red,csdlne](f)--(c3) node[lb=0,left]{\tm{1}};
\draw[red,csdlne](e)--(c3) node[lb,right,pos=.6]{\tm{2}};
\draw[red,csdlne](d)--(c4) node[lb,above,pos=.7]{\tm{2}};
\draw[red,csdlne](e)--(c4) node[lb=2pt,below,pos=.7]{\tm{1}};
\end{tikzpicture}
}%
\subfloat[distribution changes\label{f:cb3}]{
\begin{tikzpicture}
\tiny
\path[use as bounding box] (-.4,-.2) rectangle (2.9,1.9);

\draw[<->] (0.05,1.9) -- (0.05,0.1) -- (2.9,0.1);
\foreach \x in {0,1,2,3} \node(\x) at (\x*0.7+0.4,0) {$\x$};
\node[txt,rotate=90] at (-.15,.1) {\scriptsize fraction of \\ \scriptsize
content nodes};
\node[txt,anchor=center] at (1.5,-0.2) {\scriptsize triadic cardinality};

\begin{scope}[yshift=1mm]
\draw[bbar] (0.225,0) rectangle (0.375,1.47); 
\draw[bbar] (0.925,0) rectangle (1.075,0.52); 
\draw[bbar] (1.625,0) rectangle (1.775,0.05); 
\draw[bbar] (2.325,0) rectangle (2.475,0.05); 
\draw[curve,blue] plot[smooth] coordinates {(0.3,1.47) (1,0.52) (1.7,0.08)
(2.4,0.05)};
\node[blue] at (0.75,1.75) (ex1) {\scriptsize normal};
\draw[arr,blue] (ex1) -- (.5,1.35);
\node[blue] at (0.3,1.55) {$.75$};
\node[blue] at (0.98,0.6) {$.25$};
\node[blue] at (1.7,0.12) {$0$};
\node[blue] at (2.4,0.12) {$0$};

\draw[rbar]  (0.425,0) rectangle (0.575,0.05); 
\draw[rbar]  (1.125,0) rectangle (1.275,1.47); 
\draw[rbar]  (1.825,0) rectangle (1.975,0.52); 
\draw[rbar]  (2.525,0) rectangle (2.675,0.05); 
\draw[curve,red] plot[smooth] coordinates {(0.5,0.05) (1.16,1.47) (1.9,0.52)
(2.6,0.05)};
\node[red,txt] at (1.8,1.4) (ex2) {\scriptsize burst\\\scriptsize occurs};
\draw[arr,red] (ex2) -- (1.67,1);
\node[red] at (0.5,0.145) {$0$};
\node[red] at (1.2,1.55) {$.75$};
\node[red] at (1.9,0.6) {$.25$};
\node[red] at (2.6,0.12) {$0$};
\end{scope}
\end{tikzpicture}
}

  \vspace{-5pt}
  \caption{Cascading burst and influence triangle}
  \label{fig:cb_exam}
  \vspace{-8pt}
\end{figure}

\header{Influence triangles in user-content interactions.}
We say that a media content piece becomes \emph{bursty} if many users
interact with it in a short time period.
There are many reasons why a user interacts with a piece of media content.
Here, we are particularly interested in the case where one user
\emph{influences} another user to interact with the content, a.k.a.
the cascading diffusion~\cite{Leskovec2007b} or
word-of-mouth spreading~\cite{Rodrigues2011}.
It is known that many emerging news stories arising from OSNs are related
to this mechanism such as the story about the killing of Osama bin
Laden~\cite{Tsotsis2011}.
We find that a bursty media content piece formed by this mechanism is
associated with triangle formations in a network.
To illustrate this, consider Fig.~\ref{f:cb1}, in which there are five user
nodes $\{a,b,d,e,u\}$ and four content nodes $\{c_1,c_2,c_3,c_4\}$.
A directed edge between two users means that one follows another, and an
undirected edge labeled with a timestamp between a user node and a content
node represents an interaction between the user and the content at the
labeled time.
We say content node $c$ has an {\em influence triangle} if there exist two
users $a,b$ such that $a$ follows $b$ and $a$ interacts with $c$
\emph{later} than $b$ does.
In other words, the reason $a$ interacts with $c$ is due to the influence
of $b$ on $a$.
In Fig.~\ref{f:cb1}, only $c_2$ has an influence triangle, the others have
no influence triangle, meaning that the majority of user-content
interactions are not due to influence; while in Fig.~\ref{f:cb2}, every
content node is part of at least one influence triangle, meaning that many
content pieces are spreading in a cascading manner in the OSN.
From the perspective of an OSN manager who wants to know the operation
status of the OSN, if the OSN suddenly switches to a state similar to
Fig.~\ref{f:cb2} (from a previous state similar to Fig.~\ref{f:cb1}), he
knows that a {\em cascading burst} is present in the OSN.

\header{Characterizing bursts.}
So far, we find a common structure shared by different types of bursts:
the emergence of {\em interaction bursts} (caused by user-user interaction)
and {\em cascading bursts} (caused by user-content interaction) are both
accompanied with the formation of triangles, i.e., interaction or influence
triangles, in appropriately defined networks.
This finding motivates us to characterize patterns of bursts in an OSN by
characterizing the triangle statistics of a network, which we called the {\em
  triadic cardinality distribution}.

{\em Triadic cardinality} of a node in a network, e.g., a user node in
Fig.~\ref{f:ib1} or a content node in Fig.~\ref{f:cb1}, is the number of
triangles that it belongs to.
The triadic cardinality distribution then characterizes the fractions of
nodes with certain triadic cardinalities.
When a burst occurs, because many new interaction/influence triangles are
formed, we will observe that some nodes' triadic cardinalities increase,
and this results in the distribution shifting to right, as illustrated in
Figs.~\ref{f:ib3} and~\ref{f:cb3}.
The triadic cardinality distribution provides succinct summary information
to characterize burst patterns of a large scale OSN.
Hence, by tracking triadic cardinality distributions, we can detect the
presence of bursts.

In this paper, we assume that user interactions are aggregated
chronologically to form a {\em social activity stream}, which can be
considered as an abstraction of a {\em tweet stream} in Twitter, or a {\em
  news feed} in Facebook.
We aim to calculate triadic cardinality distributions from this stream.
However, when a network is large or users are very active, the social
activity stream will be high speed.
For example, the speed of the Twitter's tweets stream can be as high as
$5,700$ tweets per second on average, $143,199$ tweets per second during
the peak time, and about $500$ million to $12$ billion tweets are
aggregated per day~\cite{Krikorian2013}.
To handle such a high-speed social activity stream, we design a
sample-estimate solution, which provides a {\em maximum likelihood
  estimate} of the triadic cardinality distribution using sampled data.
Our method works in a near-real-time fashion, and is demonstrated to be
accurate and efficient.

Overall, we make three contributions in this work.
\begin{itemize}\setlength{\itemsep}{0pt}
\item We propose a useful measure, triadic cardinality distribution, which
  provides succinct summary information to characterize burst patterns of
  user interactions in a large scale OSN (Section~\ref{sec:problem}).
\item We design a sample-estimate method that is able to accurately and
  efficiently estimate triadic cardinality distributions from high-speed
  social activity streams in near-real-time
  (Sections~\ref{sec:sampling} and~\ref{sec:estimate}).
\item Extensive experiments conducted on real data demonstrate the
  usefulness of the proposed triadic cardinality distribution and
  effectiveness of our sample-estimate solution. We also show how to apply
  our method to detect bursts in Twitter during the 2014 Hong Kong Occupy
  Central movement (Section~\ref{sec:experiment}).
\end{itemize}

\section{Problem Formulation}
\label{sec:problem}

We first formally define the notion of social activity stream as mentioned
in previous section.
Then we define triadic cardinality distribution and describe our proposed
solution.

\subsection{Social Activity Stream}

We represent an OSN by $G(V,E,C)$, where $V$ is a set of users, $E$ is a
set of relationships among users, and $C$ is a set of media content such as
hashtags and video links.
Here, a relationship between two users can be undirected like the friend
relationship in Facebook, or directed like the follower relationship in
Twitter.

Users in the OSN generate {\em social activities}, e.g., interact with
other users in $V$, or content in $C$.
We denote a social activity by $a\!\in\! V\!\times\!(V\!\cup\! C)\!\times\!
[0,\infty)$.
Here user-user interaction, $a=(u,v,t)$, corresponds to user $u$
interacting with user $v$ at time $t$; and user-content interaction,
$a=(u,c,t)$, corresponds to user $u$ interacting with content $c$ at
time $t$.

These social activities are aggregated chronologically to form a {\em
social activity stream}, denoted by $S=\{a_1,a_2,\ldots\}$, where $a_k$
denotes the $k$-th social activity in the stream.

\subsection{Triadic Cardinality Distribution}

Triadic cardinality distributions are defined on  two {\em interaction
multi-graphs} which are formed by user-user and user-content interactions,
respectively.

\header{Interaction multi-graphs.}
Within a time window (e.g., an hour, a day or a week), user-user
interactions in stream $S$ form a multi-graph $\Guu(V,\Euu)$, where $V$ is
the original set of users, and $\Euu$ is a multi-set consisting of
user-user interactions in the window.
The {\em triadic cardinality of a user} $u\!\in\! V$ is the number of
interaction triangles related to $u$ in $\Guu$.
For example, user $u$ in Fig.~\ref{f:ib1} has triadic cardinality two,
and all other users have triadic cardinality one.

Similarly, user-content interactions also form a multi-graph $\Guc(V\cup
C,E\cup \Euc)$ in a time window.
Unlike $\Guu$, the node set includes both user nodes $V$ and content nodes
$C$, and the edge set includes user relations $E$ and a multi-set $\Euc$
denoting user-content interactions in the window.
Note that in $\Guc$, triadic cardinality is only defined for content nodes,
and the {\em triadic cardinality of a content node} $c\in C$ is the number of
influence triangles related to $c$ in $\Guc$.
For example, in Fig.~\ref{f:cb1}, content $c_2$ has triadic cardinality
one, and all other content nodes have triadic cardinality zero.

\header{Triadic cardinality distribution.}
Let $\theta=(\theta_0,...,\theta_W)$ and
$\vartheta=(\vartheta_0,...,\vartheta_{W'})$ denote
the triadic cardinality distributions on $\Guu$ and $\Guc$ respectively.
Here, $\theta_i$ ($\vartheta_i$) is the fraction of users (content pieces)
with triadic cardinality $i$, and $W$ ($W'$) is the maximum triadic
cardinality in $\Guu$ ($\Guc$).

The importance of the triadic cardinality distribution lies in its
capability of providing succinct summary information to characterize burst
patterns in a large scale OSN.
By tracking triadic cardinality distributions, we will discover
burst occurrences in an OSN.

\subsection{Overview of Our Solution}

\begin{figure}[htp]
\centering
\scriptsize
\begin{tikzpicture}[
txt/.style={minimum height=2em, align=center},
blk/.style={draw, rectangle, minimum height=2.3em,
  minimum width=12mm, text depth=\depthof{g}},
arr/.style={single arrow, single arrow head extend=#1,
  fill=gray!30, align=center, minimum height=8mm}
]
\node(osn) at (0,0) {\includegraphics[width=.85cm]{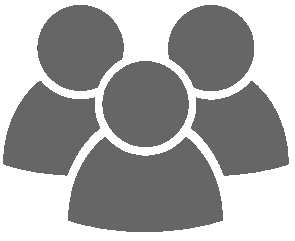}};
\node [txt,below=0 of osn] {OSN};
\node(sa) [arr=2ex,right =0 of osn] {social \\ activities};
\node(sp) [blk,right =0 of sa] {sample};
\node(a1) [arr=1ex,single arrow head extend=1ex,right =0 of sp] {};
\node(es) [blk,right =0 of a1] {estimate};
\node(a2) [arr=1ex,right =0 of es] {};
\node(gp) [draw,dashed,fit={(sp) (es)}] {};
\node [txt,below=0 of gp] {proposed solution};
\begin{scope}[shift={($(a2.east)+(.05,-.4)$)}]
  \draw[<->,thick] (0,.85)--(0,0)--(.9,0);
  \draw[very thick] (.05,.7) .. controls (.2,.2) .. (.7,.05);
  \node[txt] at (.4,-1.2em) {triadic cardinality \\ distribution};
\end{scope}
\end{tikzpicture}

\vspace{-15pt}
\caption{Overview of our sample-estimate solution}
\label{fig:solution}
\vspace{-5pt}
\end{figure}
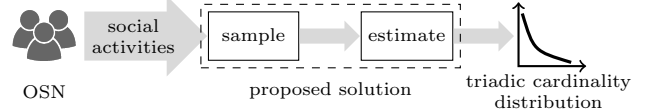

We propose an on-line solution capable of tracking the triadic cardinality
distribution from a high-speed social activity stream, as illustrated in
Fig.~\ref{fig:solution}.

Our solution consists of sampling a social activity stream in a time window
maintaining only summary statistics, and constructing an estimate of the
triadic cardinality distribution from the summary statistics at the end of
a time window.
The advantage of this approach is that it is unnecessary to store all of
the samples in the stream, and enables us to detect bursts in a
near-real-time fashion.

\section{Stream Sampling Method}
\label{sec:sampling}

In this section, we introduce the sampling method in our solution.
The purpose of sampling is to reduce the computational cost in handling the
massive amount of data in a high-speed social activity stream.

\subsection{Sampling Stream with a Coin}

The stream sampling method works as follows.
We toss a biased coin for each social activity $a\in S$.
We keep $a$ with probability $p$, and ignore it with probability $1-p$.
Hence, each social activity is independently sampled, and at the end of the
time window, only a fraction $p$ of the stream is kept.
We use these samples to obtain a summary statistics of the stream in the
current window, which we describe later.

\subsection{Probability of Sampling a Triangle}

When social activities in the stream are sampled, triangles in $\Guu$ and
$\Guc$ are sampled accordingly.
Obviously, the probability of sampling a triangle depends on $p$.
In what follows, we analyze the relationship between triangle sampling
probability and $p$, for an interaction triangle and an influence triangle,
respectively.

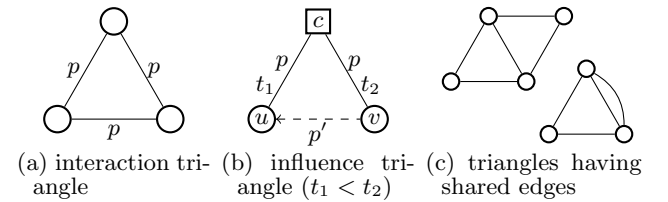
\begin{figure}[htp]
\vspace{-10pt}
\centering
\footnotesize
\tikzstyle{usr}=[draw,thick,circle,minimum size=10pt,inner sep=0]
\tikzstyle{susr}=[usr,minimum size=6.2pt]
\tikzstyle{con}=[draw,thick,rectangle,minimum size=9pt,inner sep=0]
\tikzstyle{lb}=[midway,inner sep=2pt]

\subfloat[interaction triangle\label{f:t1_pd}]{
\begin{tikzpicture}
\path[use as bounding box] (-.5,-.3) rectangle (2,1.5);
\node(u) [usr] at (0,0) {};
\node(v) [usr] at (1.5,0) {};
\node(w) [usr] at (.75,1.3) {};
\draw (u) -- (v) node[lb,below] {$p$};
\draw (u) -- (w) node[lb,left] {$p$};
\draw (v) -- (w) node[lb,right] {$p$};
\end{tikzpicture}
}%
\subfloat[influence triangle ($t_1<t_2$)\label{f:t2_pd}]{
\begin{tikzpicture}
\path[use as bounding box] (-.5,-.3) rectangle (2,1.5);
\node(u) [usr] at (0,0) {$u$};
\node(v) [usr] at (1.5,0) {$v$};
\node(c) [con] at (.75,1.3) {$c$};
\draw[<-,dashed] (u) -- (v) node[lb,below,inner sep=1pt] {$p'$};
\draw (u)--(c) node[lb,left,pos=.6] {$p$} node[lb,left,pos=.3] {$t_1$};
\draw (v)--(c) node[lb,right,pos=.6] {$p$} node[lb,right,pos=.3] {$t_2$};
\end{tikzpicture}
}%
\subfloat[triangles having shared edges\label{f:t_dependence}]{
\begin{tikzpicture}
\path[use as bounding box] (-.3,-.3) rectangle (2.5,1.5);
\begin{scope}[yshift=5mm]
\node(a) [susr] at (0,0) {};
\node(b) [susr] at (1,0) {};
\node(c) [susr] at (.5,.87) {};
\node(d) [susr] at (1.5,.87) {};
\draw (a) -- (b) -- (c) -- (a);
\draw (b) -- (d) -- (c);
\end{scope}
\begin{scope}[xshift=1.3cm,yshift=-2mm]
\node(a) [susr] at (0,0) {};
\node(b) [susr] at (1,0) {};
\node(c) [susr] at (.5,.87) {};
\draw (a) -- (b) -- (c) -- (a);
\draw (b) to [bend right] (c);
\end{scope}
\end{tikzpicture}
}

\caption{Sampling triangles.
  A solid edge represents an interaction, and a dashed edge represents a
  user relation in $E$ (i.e., a social edge). Figure (c) illustrates two
  cases of two interaction triangles having shared edges.}
\label{fig:triangle}
\vspace{-10pt}
\end{figure}

\header{Probability of sampling an interaction triangle.}
Sampling an interaction triangle, which consists of three user-user
interaction edges in $\Euu$, is equivalent to all its three edges being
sampled.
Because each interaction edge is independently sampled with probability
$p$, then an interaction triangle is sampled with probability
$p^3$, as illustrated in Fig.~\ref{f:t1_pd}.

\header{Probability of sampling an influence triangle.}
Calculating the probability of sampling an influence triangle is more
complicated.
First, we know that an influence triangle consists of two user-content
interaction edges in $\Euc$ and one social edge in $E$.
Second, we note that stream sampling only applies to edges in $\Euc\cup
\Euu$; edges in $E$ are not sampled as they do not appear in the social
activity stream.

In Fig.~\ref{f:t2_pd}, suppose we have sampled two user-content interaction
edges $uc$ and $vc$, and assume user $u$ interacted with content $c$
earlier than user $v$.
To determine whether content $c$ has an influence triangle formed by $u$
and $v$, we need to check whether (directed) edge $(v,u)$ exists in $E$.
This can be done by querying neighbors of one of the two users in the OSN.
For example, in Twitter, we query \emph{followees} of $v$ and check whether
$v$ follows $u$; or in Facebook, we query friends of $v$ and check whether
$u$ is a friend of $v$.

Suppose we observe $n_c$ sampled users that all interact with $c$ during
the current time window, denoted by $V_c=\{u_1,\ldots,u_{n_c}\}$ where
$u_i$ interacted with $c$ earlier than $u_j$, $i<j$.
To verify every sampled triangle related to $c$, we need to query the OSN
$n_c(n_c-1)/2$ times.
This query cost is obviously expensive when $n_c$ is large.
To reduce this query cost, instead of checking every possible user pair, we
check a user pair with probability $p'$.
This is equivalent to sampling a social edge in $E$ with probability $p'$,
conditioned on the two associated user-content interactions having been
sampled.
Then, it is easy to see that an influence triangle is sampled with
probability $p^2p'$.

We summarize the above discussions in Theorem~\ref{th:pd}.

\begin{theorem}\label{th:pd}
  If we independently sample each social activity in stream $S$ with
  probability $p$, and check the existence of a user relation in the OSN
  with probability $p'$, then each interaction (influence) triangle in
  graph $\Guu$ ($\Guc$) is sampled with identical probability
\begin{equation}
p_\delta =
 \begin{cases}
  p^3   & \text{for an interaction triangle},\\
  p^2p' & \text{for an influence triangle}.
 \end{cases}
\label{eq:pdelta}
\end{equation}
\end{theorem}

\header{Remark.}
Although triangles of the same type are sampled identically, they may {\em
  not} be sampled {\em independently}, such as the cases two triangles have
shared edges in Fig.~\ref{f:t_dependence}.
We will consider this issue in detail in Section~\ref{sec:estimate}.

\subsection{Statistics of Sampled Data}

The above sampling process is equivalent to sampling edges in multi-graphs
$\Guu$ and $\Guc$: an activity edge $e\in\Euu\cup\Euc$ is independently
sampled with probability $p$; a social edge $e'\in E$ is sampled with
conditional probability $p'$.

At the end of the time window, we obtain two {\em sampled multi-graphs}
$\Guu'$ and $\Guc'$\footnote{In $\Guc'$, each sampled social edge $e'$
  needs to be marked with the influence triangle which $e'$ belongs to,
  corresponding to the two user-content interactions that $e'$ is checked for.}.
Calculating the triadic cardinalities for nodes in these reduced graphs
is much easier than on the original unsampled graphs.
For $\Guu'$, we calculate triadic cardinality for
each user node, and obtain statistics $g=(g_0,\ldots,g_M)$, where $g_j$,
$0\leq j\leq M$, denotes the number of nodes with $j$ triangles in $\Guu'$.
Similar statistics are also obtained from $\Guc'$, denoted by
$f=(f_0,\ldots,f_{M'})$ (where $f_j$ is the number of content nodes with
$j$ influence triangles in $\Guc'$).
We only need to store $g$ and $f$ in computer memory and use them to
estimate $\theta$ and $\vartheta$ in the next section.

\section{Estimation Methods}
\label{sec:estimate}

We are now ready to derive a maximum likelihood estimate (MLE) of the
triadic cardinality distribution using statistics obtained in the
sampling step.
The estimation in this section can be viewed as an analog of network flow size
distribution estimation~\cite{Duffield2003,Ribeiro2006} in which a packet
in a flow is viewed to be a triangle of a node.
However, in our case, triangle samples are not independent, and a node may
have no triangles.
These issues complicate estimation, and we will describe how to solve these
issues in this section.

Note that we only discuss how to obtain the MLE of $\theta$ using $g$, as
the MLE of $\vartheta$ using $f$ is easily obtained using a similar
approach.
To estimate $\theta$, we first consider the easier case where graph size
$|V|=n$ is known.
Later, we extend our analysis to the case where $|V|$ is unknown.

\subsection{MLE when Graph Size is Known}

Recall that $g_j$, $0\leq j\leq M$, is the number of nodes with $j$ sampled
triangles in $\Guu'$.
First, note that observing a node with $j$ sampled triangles in $\Guu'$
implies that the node has at least $j$ triangles in $\Guu$.

We also need to pay special attention to $g_0$, which is the number of
nodes with no triangle in $\Guu'$.
Due to sampling, some nodes may be unobserved (e.g., no edge attached
to the node is sampled), and these unobserved nodes also have no sampled
triangle.
We include these in $g_0$; the advantage of this inclusion will be seen
later.
Since we have assumed a total of $n$ nodes in $\Guu$, the number of
unobserved nodes is $n-\sum_{j=0}^M g_j$.
Therefore, we calibrate $g_0$ by
\[
g_0 \triangleq n-\sum_{j=1}^M g_j.
\]

Our goal is to derive an MLE of $\theta$.
To this end, we need to model the sampling process.
For a randomly chosen node, let $X$ denote the number of triangles to which
it belongs in $\Guu$, and let $Y$ denote the number of triangles observed
during sampling.
Then $P(Y=j|X=i), 0\leq j\leq i$, is the conditional probability that a
node has $j$ sampled triangles in $\Guu'$ given that it has $i$ triangles
in $\Guu$.
The sampling of a triangle can be viewed as a Bernoulli trial with
a success probability of $p_\delta$, according to Theorem~\ref{th:pd}.
If Bernoulli trials are independent, which means triangles are
independently sampled, then $P(Y=j|X=i)$ follows a binomial distribution.
However, independence does not hold for triangles having shared edges, as
illustrated in Fig.~\ref{f:t_dependence}.
As a result, it is non-trivial to derive $P(Y=j|X=i)$ with the
existence of dependence.
To deal with this dependence, we approximate sums of dependent Bernoulli
random variables by a Beta-binomial distribution~\cite{Yu2002}, which yields
\begin{align*}
& \hspace{-0.3in} P(Y=j|X=i)
= BetaBin(j|i,p_\delta/\alpha,(1-p_\delta)/\alpha) \\
=& \binom{i}{j}
\frac{\prod_{s=0}^{j-1}(s\alpha+p_\delta)\prod_{s=0}^{i-j-1}(s\alpha+1-p_\delta)}
{\prod_{s=0}^{i-1}(s\alpha+1)}
\triangleq b_{ji}(\alpha)
\end{align*}
where $\prod_0^{-1}\triangleq 1$.
The above Beta-binomial distribution parameterized by $\alpha$ allows
pairwise identically distributed Bernoulli trials to have covariance
$\alpha p_\delta(1-p_\delta)/(1+\alpha)$.
It reduces to a binomial distribution when $\alpha=0$.
We have carried out $\chi^2$ goodness-of-fit tests and the results
demonstrate that the above model indeed fits well the observed data on many
graphs (and is always better than the binomial model, of course).

Using this model, we easily obtain the likelihood of observing a node
to have $j$ sampled triangles, i.e.,
\[
P(Y=j)
= \sum_{i=j}^W P(Y=j|X=i)P(X=i)
= \sum_{i=j}^W b_{ji}(\alpha)\theta_i.
\]
Then, the log-likelihood of all observations $\{Y_k=y_k\}_{k=1}^n$, where
$Y_k=y_k$ denotes the $k$-th node having $y_k$ sampled triangles, yields
\begin{equation}
\CL(\theta,\alpha)
\!\triangleq\!\log P(\{Y_k\!=\!y_k\}_{k=1}^n)
\!=\!\sum_{j=0}^M g_j \log\!\sum_{i=j}^W b_{ji}(\alpha)\theta_i.
\label{eq:log-likelihood}
\end{equation}

The MLE of $\theta$ can then be obtained by
maximizing~\eqref{eq:log-likelihood} with respect to $\theta$ and $\alpha$
under the constraint that $\sum_{i=0}^W\theta_i=1$.
Note that this is non-trivial due to the summation inside the $\log$
operation.
In the next subsection, we use the expectation-maximization (EM) algorithm
to obtain the MLE in a more convenient way.

\subsection{EM Algorithm when Graph Size is Known}

If we already know that the $k$-th node has $x_k$ triangles in $\Guu$,
i.e., $X_k=x_k$, then the complete likelihood of observations
$\{(Y_k,X_k)\}_{k=1}^n$ is
\begin{align*}
& P(\{(Y_k,X_k)\}_{k=1}^n)
=\prod_{k=1}^n P(Y_k=y_k, X_k=x_k) \\
&=\prod_{j=0}^M \prod_{i=j}^W P(Y=j, X=i)^{z_{ij}}
=\prod_{j=0}^M \prod_{i=j}^W \left[b_{ji}(\alpha)\theta_i\right]^{z_{ij}}
\end{align*}
where $z_{ij}=\sum_{k=1}^n \indr{x_k=i\wedge y_k=j}$ is the number of
nodes with $i$ triangles and $j$ of them being sampled
(and $\indr{\cdot}$ is the indicator function).
The complete log-likelihood is
\begin{equation}
\CL_c(\theta,\alpha)
\triangleq\sum_{j=0}^M \sum_{i=j}^W z_{ij}\log\left[ b_{ji}(\alpha)\theta_i \right].
\label{eq:c_log-likelihood}
\end{equation}
Here, we can treat $\{X_k\}_{k=1}^n$ as hidden variables, and apply the EM
algorithm to calculate the MLE.

\header{E-step:}
We calculate the expectation of the complete
log-likelihood in Eq.~\eqref{eq:c_log-likelihood} with respect to hidden
variables $\{X_k\}_k$, conditioned on data $\{Y_k\}_k$ and
previous estimates $\theta^{(t)}$ and $\alpha^{(t)}$. That is
\[
Q(\theta,\alpha;\theta^{(t)},\alpha^{(t)})
\triangleq \sum_{j=0}^M \sum_{i=j}^W \E[\theta^{(t)},\alpha^{(t)}]{z_{ij}}
\log\left[ b_{ji}(\alpha)\theta_i \right].
\]
Here, $\E[\theta^{(t)},\alpha^{(t)}]{z_{ij}}$ can be viewed as the average
number of nodes that have $i$ triangles in $\Guu$, of which $j$ are
sampled.
Because
\begin{align*}
 & \hspace{-0.5in} P(X=i|Y=j,\theta^{(t)},\alpha^{(t)}) \\
=&\, \frac{P(Y=j|X=i,\alpha^{(t)})P(X=i|\theta^{(t)})}
   {\sum_{i'}P(Y=j|X=i',\alpha^{(t)})P(X=i'|\theta^{(t)})} \\
=&\, \frac{b_{ji}(\alpha^{(t)})\theta_i^{(t)}}
   {\sum_{i'} b_{ji'}(\alpha^{(t)})\theta_{i'}^{(t)}}
\triangleq p_{i|j}
\end{align*}
and we have observed $g_j$ nodes with $j$ sampled triangles,
then $\E[\theta^{(t)},\alpha^{(t)}]{z_{ij}} = g_jp_{i|j}$.

\header{M-step:}
We now maximize $Q(\theta,\alpha;\theta^{(t)},\alpha^{(t)})$ with respect to
$\theta$ and $\alpha$ subject to the constraint $\sum_{i=0}^W \theta_i=1$.
After the $\log$ operation, $\theta$ and $\alpha$ are well separated.
Hence, we obtain
\begin{align*}
\theta_i^{(t+1)}
&= \arg\max_\theta Q(\theta,\alpha;\theta^{(t)},\alpha^{(t)}) \\
&= \frac{\sum_{j=0}^i \E[\theta^{(t)},\alpha^{(t)}]{z_{ij}}}
   {\sum_{j=0}^M \sum_{i'=j}^W \E[\theta^{(t)},\alpha^{(t)}]{z_{i'j}}},
   \quad 0\leq i\leq W,
\end{align*}
and
$\alpha^{(t+1)}=\arg\max_\alpha Q(\theta,\alpha;\theta^{(t)},\alpha^{(t)})$,
which can be solved using gradient descent methods.

Multiple iterations of the E-step and the M-step, EM algorithm converges to
a solution, which is a local maximum of~\eqref{eq:log-likelihood}.
We denote this solution by $\hat\theta$ and $\hat\alpha$.

\subsection{MLE when Graph Size is Unknown}
\label{subsec:mle_n_unknown}

When the graph size is unknown, one can use probabilistic counting methods
such as loglog counting~\cite{Durand2003} to obtain an estimate of graph
size from the stream, and then apply our previously developed method to
obtain estimate $\hat\theta$.
Note that this introduces additional statistical errors to $\hat\theta$ due
to the inaccurate estimate of the graph size.
In what follows, we slightly reformulate the problem and develop a method
that can simultaneously estimate both the graph size and the triadic
cardinality distribution from the sampled data.

When the graph size is unknown, we cannot calibrate $g_0$ because we
do not know the number of unsampled nodes.
A node of degree $d$ is not sampled with probability $(1-p)^d$.
There is no clear relationship between an unsampled node and its triadic
cardinality.
As a result, we cannot easily model the absence of nodes by
$\theta$, and this complicates estimation design.

To solve this issue, we need to slightly reformulate our problem: (i)
instead of estimating the total number of nodes in $\Guu$, we estimate the
number of nodes belonging to at least one triangle in $\Guu$, denoted by $n_+$;
(ii) we estimate the triadic cardinality distribution
$\theta^+=(\theta_1^+,\ldots,\theta_W^+)$, where $\theta_i^+$ is the
fraction of nodes with $i$ triangles over the nodes having at least one
triangle in $\Guu$.

\header{Estimating $n_+$.}
Under the Beta-binomial model, the probability that a node has $i$
triangles in $\Guu$, of which none are sampled, is
\[
q_i(\alpha)
\triangleq P(Y=0|X=i)
= \prod_{s=0}^{i-1}\left(1-\frac{p_\delta}{s\alpha+1}\right).
\label{eq:qi}
\]
Then, the probability that a node has triangles in $\Guu$, of which none
are sampled, is
\[
q(\theta^+,\alpha)
\triangleq P(Y=0|X\geq 1)
= \sum_{i=1}^W q_i(\alpha)\theta_i^+.
\]
Because there are $\sum_{j=1}^M g_j$ nodes having been observed to have
at least one sampled triangle.
Hence, $n_+$ can be estimated by
\begin{equation}
\hat{n}_+ = \frac{\sum_{j=1}^M g_j}{1-q(\theta^+,\alpha)}.
\label{eq:nplus}
\end{equation}

Note that estimator~\eqref{eq:nplus} relies on
$\theta^+$ and $\alpha$, and we can obtain them using the following
procedure.

\header{Estimating $\theta^+$ and $\alpha$.}
We discard $g_0$ and only use $g^+\triangleq (g_1,\ldots,g_M)$ to estimate
$\theta^+$ and $\alpha$.
The basic idea is to derive the likelihood for nodes that are observed
to have at least one sampled triangle, i.e., $\{Y_k=y_k\colon y_k\geq 1\}$.
In this case, the probability that a node has $X=i$ triangles, and $Y=j$ of
them are sampled, conditioned on $Y\geq 1$, is
\begin{align*}
&\hspace{-0.2in}P(Y=j|X=i,Y\geq 1) \\
=&\, \frac{BetaBin(j|i,p_\delta/\alpha,(1-p_\delta)/\alpha)}
{1-BetaBin(0|i,p_\delta/\alpha,(1-p_\delta)/\alpha)}
\triangleq a_{ji}(\alpha),\,\, j\geq 1.
\end{align*}
Then the probability that a node is observed to have $j$ sampled triangles,
conditioned on $Y\geq 1$, is
\begin{align*}
&\hspace{-0.15in} P(Y=j|Y\geq 1) \\
=&\, \sum_{i=j}^W P(Y\!=\!j|X\!=\!i,Y\!\geq\! 1)P(X\!=\!i|Y\!\geq\! 1)
\!=\! \sum_{i=j}^W a_{ji}(\alpha)\phi_i,
\end{align*}
where
\begin{equation}
\phi_i \triangleq P(X=i|Y\geq 1)
= \frac{\theta_i^+[1-q_i(\alpha)]}
{\sum_{i'=1}^W \theta_{i'}^+[1-q_{i'}(\alpha)]},\,\, i\geq 1.
\label{eq:phi}
\end{equation}
Now it is straightforward to obtain the previously mentioned likelihood.
Furthermore, we can leverage our previously developed EM algorithm by
replacing $\theta_i$ by $\phi_i$, $b_{ji}$ by $a_{ji}$, to obtain MLEs
for $\phi$ and $\alpha$.
We omit these details, and directly provide the final EM iterations:
\[
\phi_i^{(t+1)} =
\frac{\sum_{j=1}^i \E[\phi^{(t)},\alpha^{(t)}]{z_{ij}}}
{\sum_{j=1}^M \sum_{i'=j}^W
\E[\phi^{(t)},\alpha^{(t)}]{z_{i'j}}},\quad i\geq 1,
\]
where
\[
\E[\phi^{(t)},\alpha^{(t)}]{z_{ij}} =
\frac{g_ja_{ji}(\alpha^{(t)})\phi_i^{(t)}}
{\sum_{i'=j}^W a_{ji'}(\alpha^{(t)})\phi_{i'}^{(t)}},
\quad i\geq j\geq 1,
\]
and $\alpha^{(t+1)}=\arg\max_\alpha Q(\phi,\alpha;\phi^{(t)},\alpha^{(t)})$
is solved using gradient decent methods.

Once EM converges, we obtain estimates $\hat\phi$ and $\hat\alpha$.
The estimate for $\theta^+$ is then obtained by Eq.~\eqref{eq:phi}, i.e.,
\begin{equation}
\hat\theta_i^+ =
\frac{\hat\phi_i/[1-q_i(\hat\alpha)]}
{\sum_{i'=1}^W\hat\phi_{i'}/[1-q_{i'}(\hat\alpha)]},
\quad 1\leq i\leq W.
\label{eq:rescaling}
\end{equation}
Finally, $\hat{n}_+$ is obtained by the estimator in Eq.~\eqref{eq:nplus}.

\section{Experiments}
\label{sec:experiment}

In this section, we first empirically verify the claims we have made.
Then, we validate the proposed estimation methods on several real-world
networks.
Finally, we illustrate our method to detect bursts in Twitter during the
2014 Hong Kong Occupy Central movement.
\subsection{Analyzing Bursts in Enron Dataset}

In the first experiment, we use a public email communication dataset to
empirically show how bursts in networks can change the triadic cardinality
distribution, and verify our claims previously made.

\header{Enron email dataset.}
The Enron email dataset~\cite{Klimt2004} includes the entire email
communications (e.g., who sent an email to whom at what time) of the Enron
corporation from its startup to bankruptcy.
The used dataset is carefully cleaned by removing spamming accounts/emails
and emails with incorrect timestamps.
The cleaned dataset contains $22,477$ email accounts and $164,081$
email communications between Jan 2001 and Apr 2002.
We use this dataset to study patterns of bursts caused by email
communications among people, i.e., by user-user interactions.

\begin{figure}
\centering
\includegraphics[width=.9\linewidth]{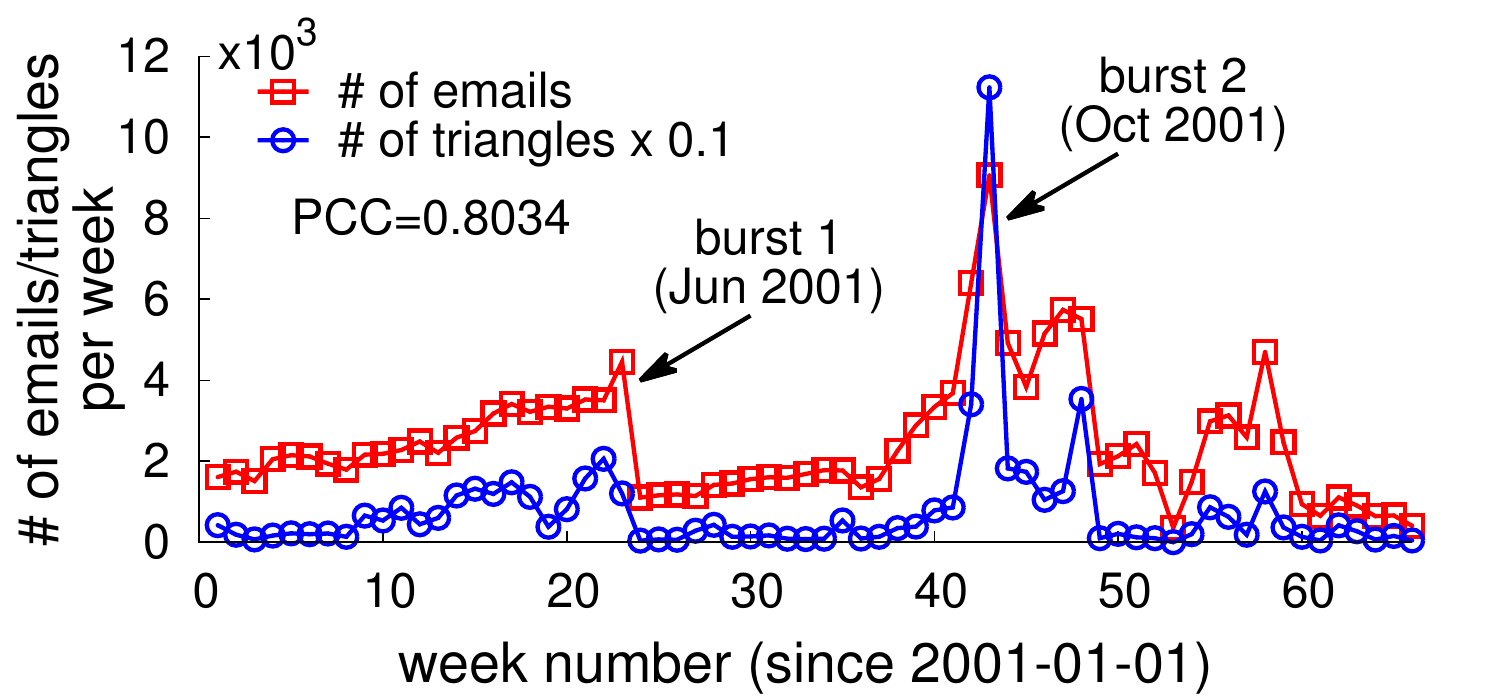}
\vspace{-10pt}
\caption{Email and triangle volumes per week\label{fig:email_volume}}
\vspace{-10pt}
\end{figure}

\header{Observations from data.}
Because the data has been cleaned, the number of user-user interactions
(i.e., number of sent emails\footnote{If an email has $x$ recipients, we
  count it $x$ times.}) per time window reliably indicates burst
occurrences.
We show the number of emails sent per week in Fig.~\ref{fig:email_volume},
and observe at least two bursts that occurred in Jun and Oct 2001,
respectively.
We also show the number of interaction triangles formed during each week.
The Pearson correlation coefficient (PCC) between the email and triangle
volum series is $0.8$, which reflects a very strong correlation.
The sudden increase (or decrease) of email volumes during the two bursts is
accompanied with the sudden increase (or decrease) of the number of
triangles.
Thus, this observation verifies our claim that the emergence
of a burst is accompanied with the formation of triangles in networks.

\begin{figure}[htp]
\vspace{-12pt}
\centering
\subfloat[Burst 1 shifts the distribution to left.
  \label{fig:burst1}]{\includegraphics[width=.5\linewidth]{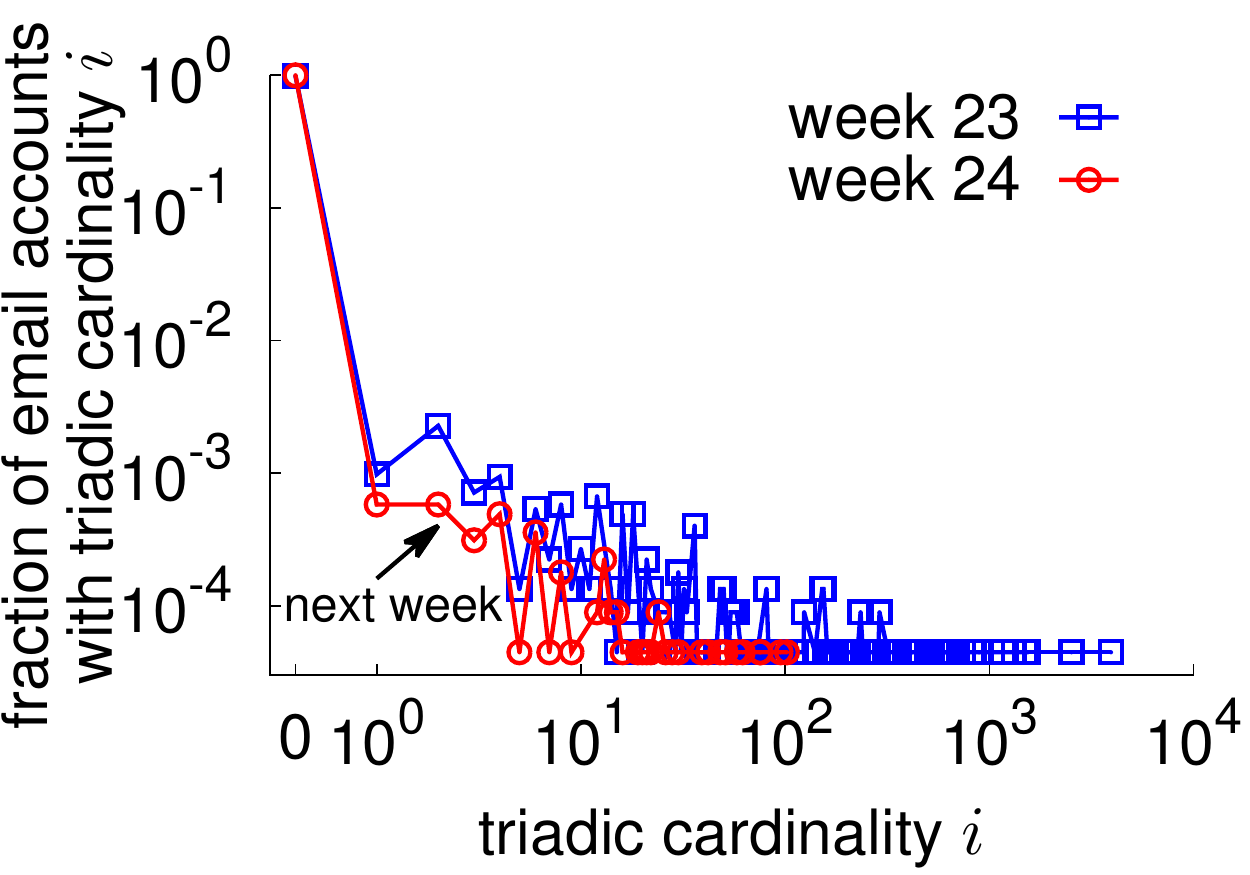}}
\subfloat[Burst 2 shifts the distribution to right.
  \label{fig:burst2}]{\includegraphics[width=.5\linewidth]{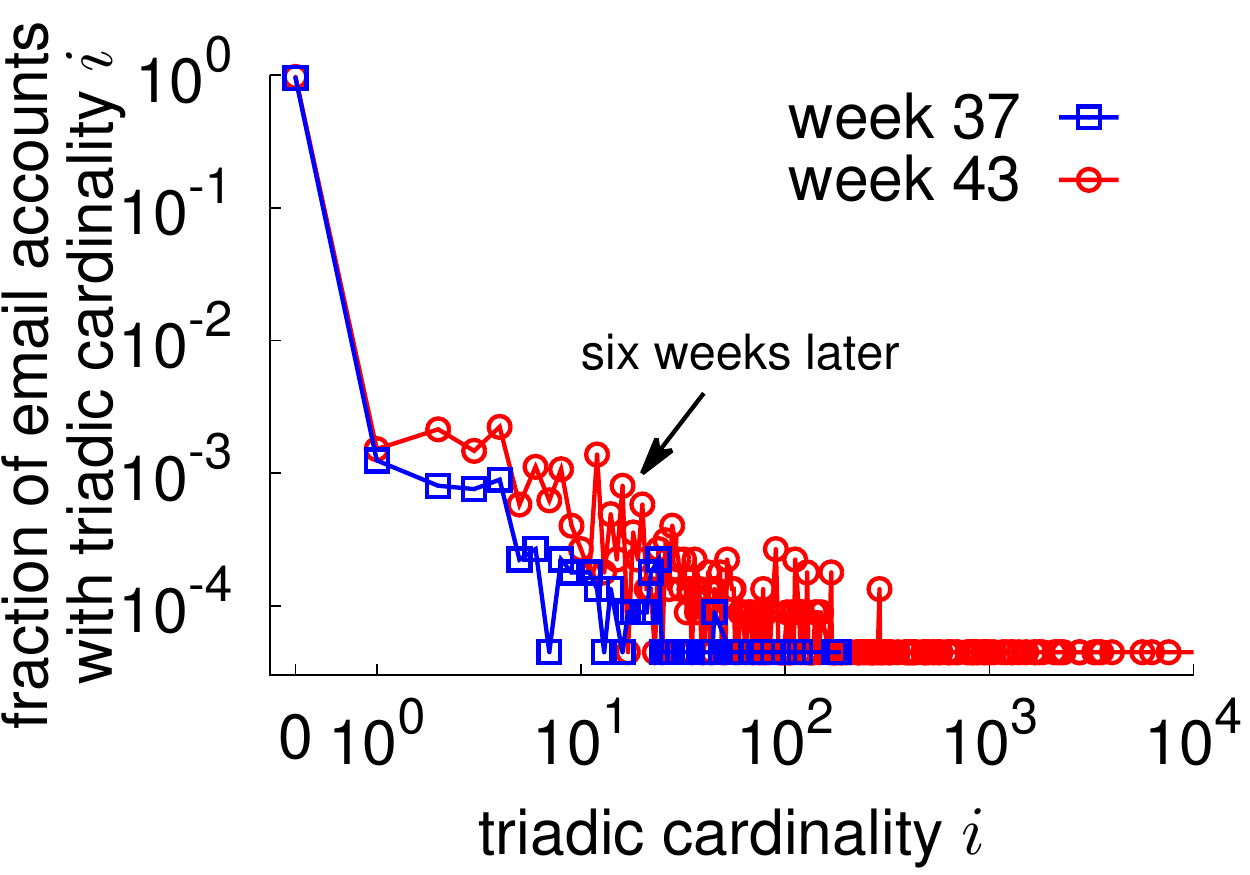}}
\vspace{-5pt}
\caption{Bursts change  distribution curves.\label{fig:burst}}
\vspace{-5pt}
\end{figure}

\header{How bursts change triadic cardinality distributions.}
Our burst detection method relies on a claim that, when a burst occurs,
the triadic cardinality distribution changes.
To see this, we show the triadic cardinality distributions before and
during the bursts in Fig.~\ref{fig:burst}.
For the first burst, due to the sudden decrease of email communications
from week 23 to week 24, we observe in Fig.~\ref{fig:burst1} that the
distribution shifts to the left.
While for the second burst, due to the gradual increase of email
communications, we observe in Fig.~\ref{fig:burst2} that the distribution
in week 43 shifts to the right in comparison to previous weeks.
Again, the observation verifies our claim that triadic cardinality
distribution changes when a burst occurs.

\header{Impacts of spam.}
As we mentioned earlier, if spam exists, simply using the volume of user
interactions to detect bursts will result in false alarms, while the triadic
cardinality distribution is a good indicator immune to spam.
To demonstrate this claim, suppose a spammer suddenly becomes active in
week 23, and generates email spams to distort the original triadic
cardinality distribution of week 23.
We consider the following two spamming strategies:
\begin{itemize}\setlength{\itemsep}{-2pt}
\item \emph{Random}: The spammer randomly chooses many target users to send
  spam.
\item \emph{Random-Friend}: At each step, the spammer randomly chooses a
  user and a random friend of the user\footnote{We assume two Enron users are friends
    if they have at least one email communication in the dataset.}, as two
  targets; and sends spams to each of these two targets. The spammer
  repeats this step a number of times.
\end{itemize}

\begin{figure}[htp]
\vspace{-15pt}
\centering
\subfloat[KL divergence\label{fig:kl}]{
    \includegraphics[width=.5\linewidth]{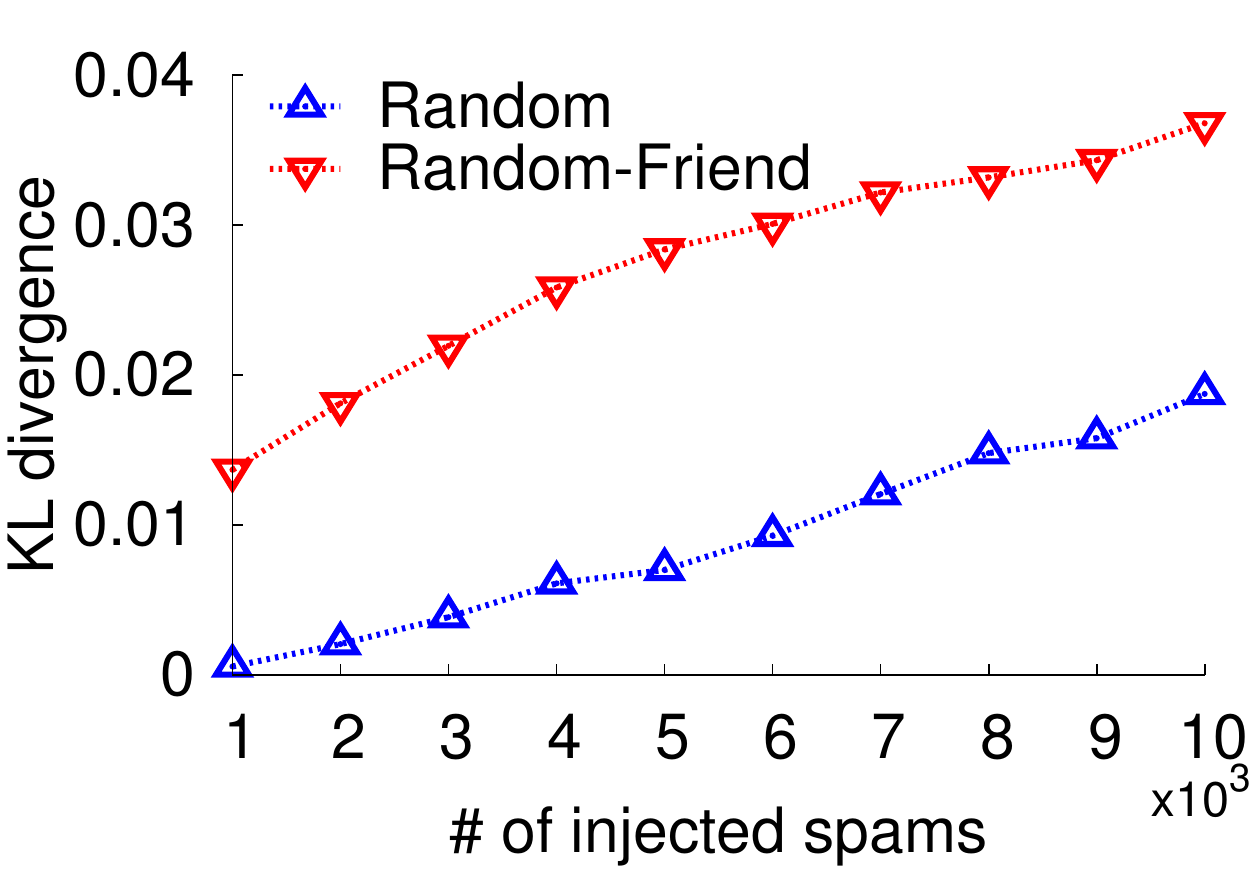}}
\subfloat[Distorted distributions\label{fig:distorted}]{
    \includegraphics[width=.5\linewidth]{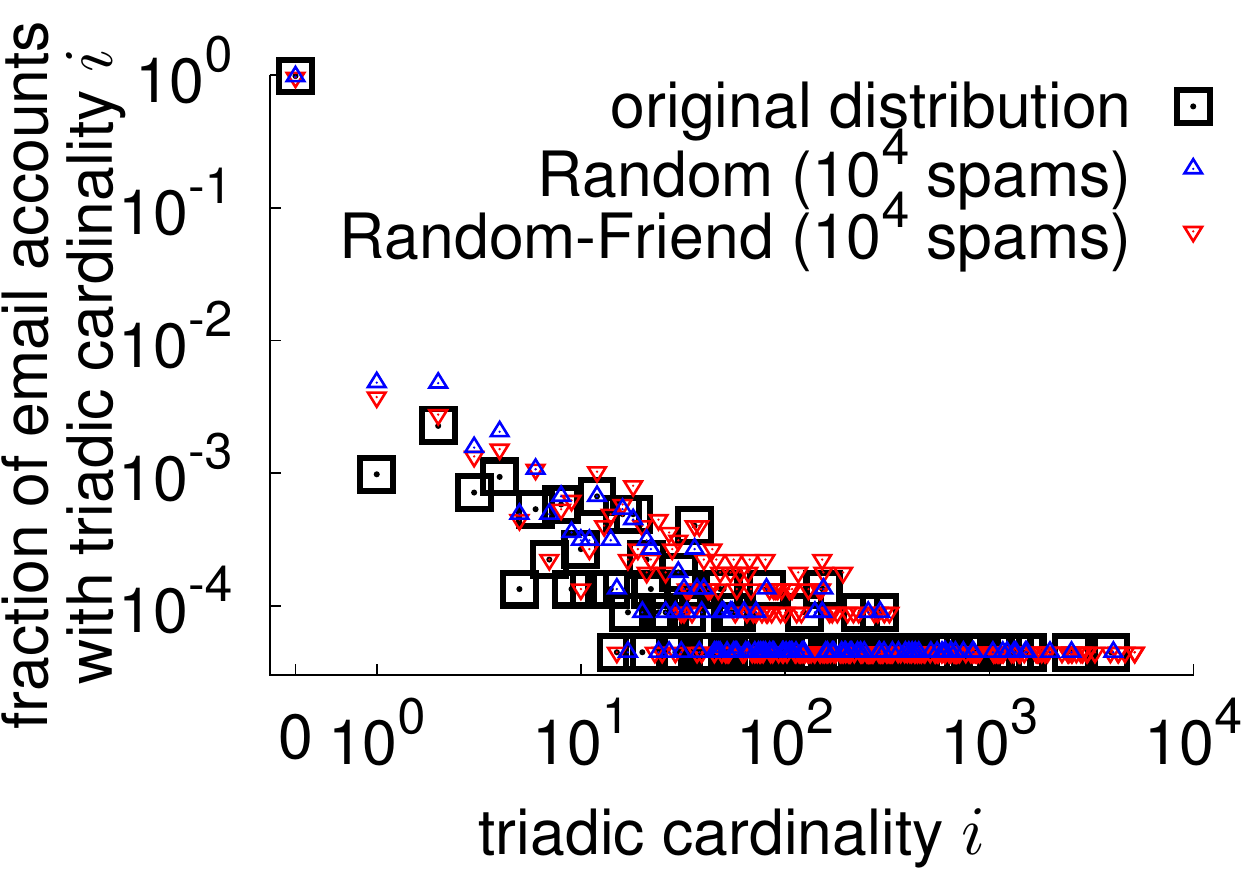}}
\vspace{-5pt}
\caption{Impacts of spam.}
\vspace{-5pt}
\end{figure}

In order to measure the extent that spams can distort the original triadic
cardinality distribution of week 23, we use Kullback-Leibler (KL)
divergence to measure the difference between the original and distorted
distributions.
The relationship between KL divergence and the number of injected spams is
shown in Fig.~\ref{fig:kl}.
For both strategies, KL divergences both increase as more spams are
injected into the interaction network, which is expected.
The Random-Friend strategy can cause larger divergences than the Random
strategy, as Random-Friend strategy is easier to introduce new triangles to
the interaction network of week 23 for the reason that two friends are
more likely to communicate in a week.
However, even when $10^4$ spams are injected, the spams incur an increasing
KL divergence of less than $0.04$.
From Fig.~\ref{fig:distorted}, we can see that the divergence is indeed
small.
(This may be explained by the ``{\em center of attention}''
phenomenon~\cite{Backstrom2011}, i.e., a person may have hundreds of
friends but he usually only interacts with a small fraction of them in a
time window. Hence, Random-Friend strategy does not form many triangles.)
Therefore, these observations verify that triadic cardinality
distribution is robust against common spamming attacks.

\subsection{Validating Estimation Methods}

In the second experiment, we demonstrate that our proposed estimation
methods produce good estimates of triadic cardinality distributions
using sampled data while reducing computational cost.

\header{Datasets.}
Because the input of our estimation methods is in fact a sampled graph, we
use several public available graphs of different types and scales from the
SNAP graph repository (\url{snap.stanford.edu/data}) as our testbeds.
We summarize statistics of these graphs in Table~\ref{tab:graphs}.

\begin{table}[htp]
\vspace{-10pt}
\centering
\caption{Network statistics\label{tab:graphs}}
\setlength{\tabcolsep}{3pt}
\begin{tabular}{l|l|r|r}
\hline\hline
Network & Type                 & Nodes       & Edges        \\
\hline
HepTh   & directed, citation   & $27,770$    & $352,807$    \\
DBLP    & undirected, coauthor & $317,080$   & $1,049,866$  \\
YouTube & undirected, OSN      & $1,134,890$ & $2,987,624$  \\
Pokec   & directed, OSN        & $1,632,803$ & $30,622,564$ \\
\hline
\end{tabular}
\vspace{-5pt}
\end{table}

For each graph, we sample an edge with probability $p$, and obtain a sampled
graph.
We then calculate the triadic cardinality for each node in the sampled
graph, and obtain statistics $g$.
Note that the estimator uses $g$ to obtain an estimate of the triadic
cardinality distribution for each graph, which is then compared with the
ground truth distribution, i.e., the triadic cardinality distribution of
the original unsampled graph, to evaluate the performance of the estimation
method.

\header{Validation when graph size is known.}
We first evaluate the estimation method when the graph size is known in
advance, as is the assumption of our first method.
The first method outputs estimate
$\hat\theta=(\hat\theta_0,\ldots,\hat\theta_W)$.

\begin{figure}
\vspace{-10pt}
\centering
\subfloat[HepTh]{\includegraphics[width=.5\linewidth]{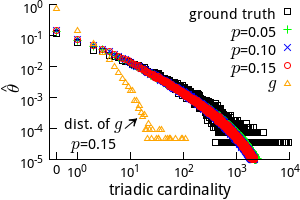}}
\subfloat[DBLP]{\includegraphics[width=.5\linewidth]{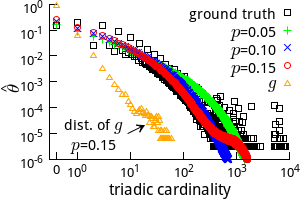}}\\[-8pt]
\subfloat[YouTube]{\includegraphics[width=.5\linewidth]{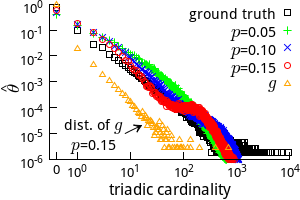}}
\subfloat[Pokec]{\includegraphics[width=.5\linewidth]{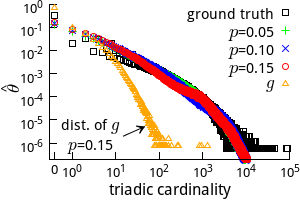}}
\vspace{-8pt}
\caption{Estimates of $\theta$ when graph size is known. $\hat\alpha$
  corresponding to each graph and $p$ is typically small, ranging from $0.00015$
  to $0.028$. $W\!=\!10^4$ and each result is averaged over $100$ runs.}
\label{fig:est1}
\vspace{-15pt}
\end{figure}

The estimates on the four graphs and comparisons with ground truth
distributions are depicted in Fig.~\ref{fig:est1}.
For each graph, we set $p=0.05,0.1$ and $0.15$, respectively.
From these results, we show that when more data is sampled the estimate
generally improves, but even when $p=0.05$ is sufficient to obtain a good
estimate.
The sampled triadic cardinality distribution of $g$ for $p=0.15$ is also
shown for each graph.
It is clear to see that the estimator has the ability to ``correct'' this
distribution to approach the ground truth distribution.

\begin{figure}[hb]
\vspace{-20pt}
\centering
\subfloat[HepTh]{\includegraphics[width=.5\linewidth]{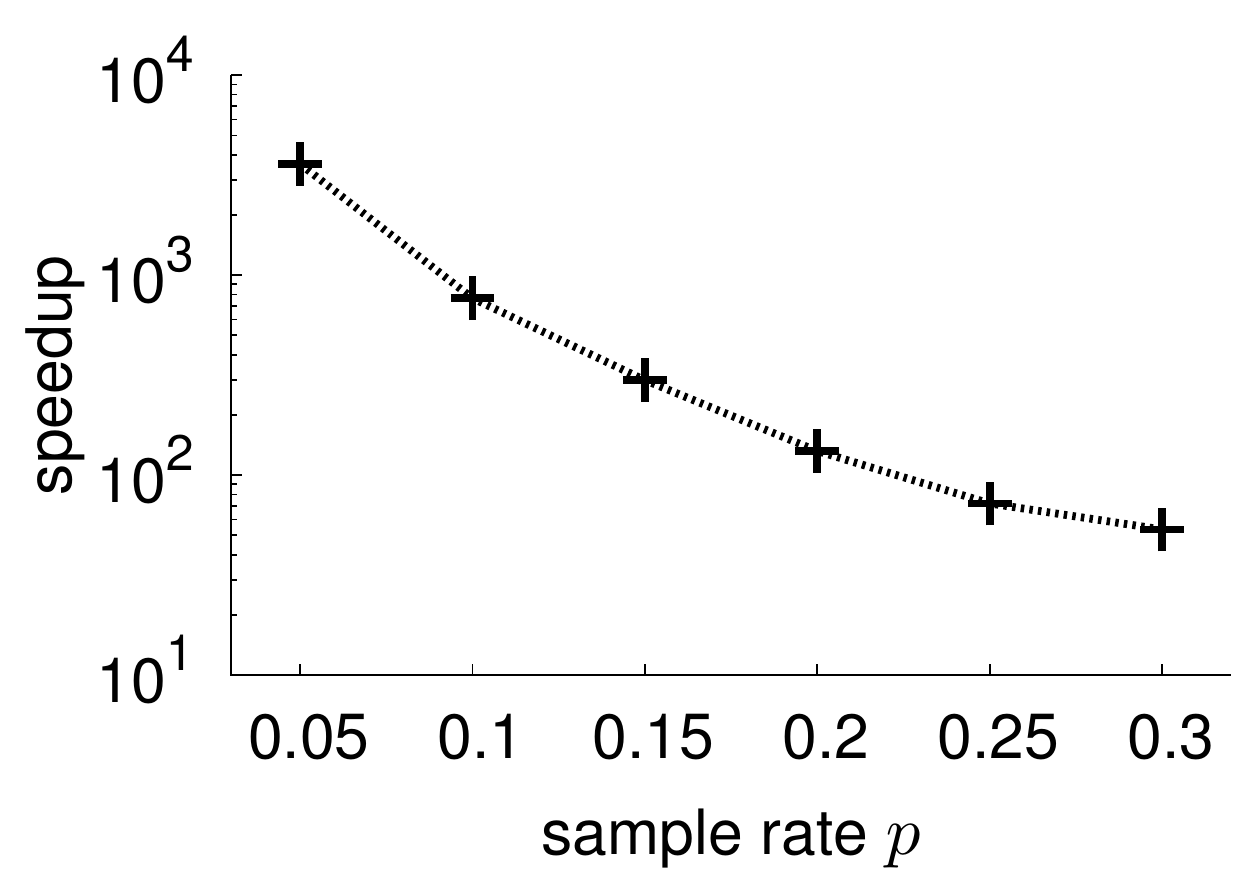}}
\subfloat[DBLP]{\includegraphics[width=.5\linewidth]{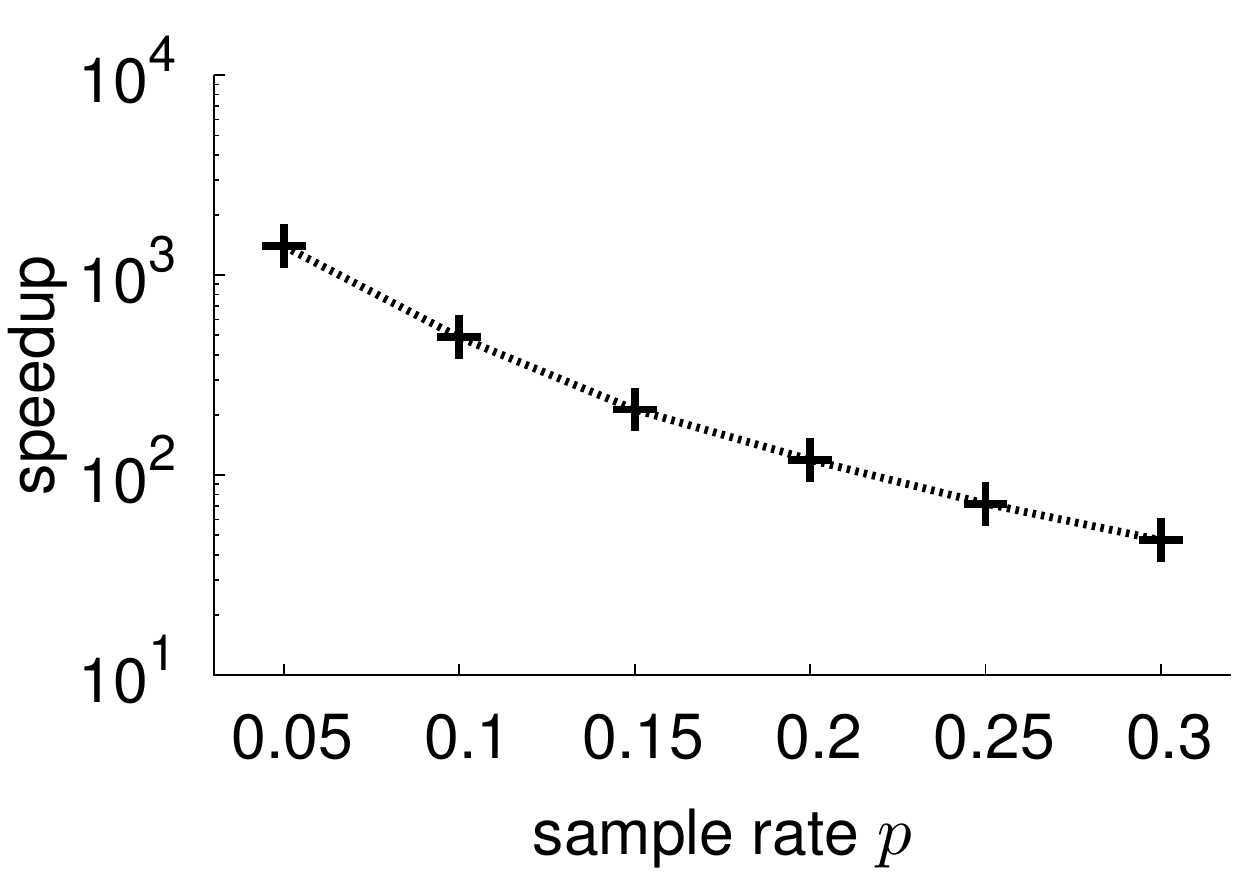}}\\[-10pt]
\subfloat[YouTube]{\includegraphics[width=.5\linewidth]{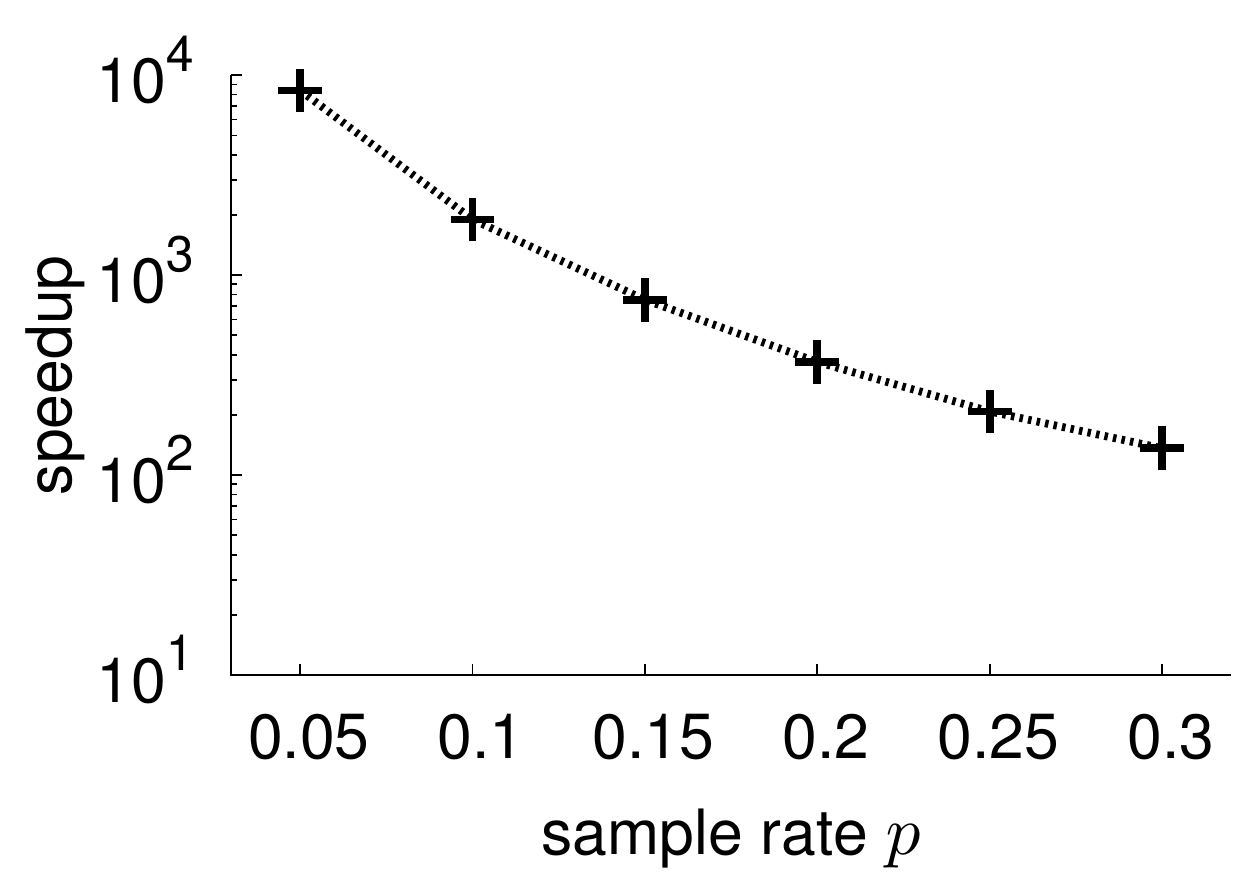}}
\subfloat[Pokec]{\includegraphics[width=.5\linewidth]{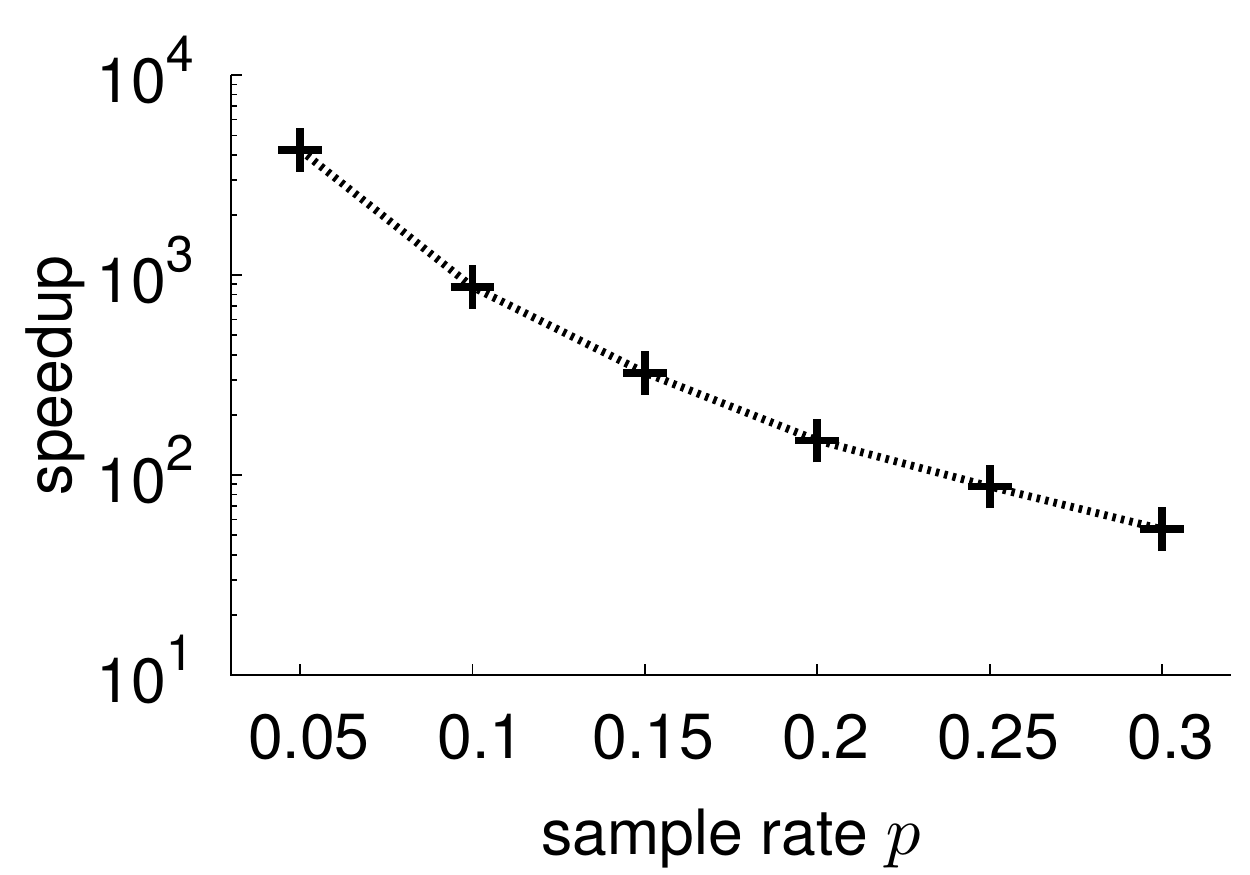}}
\vspace{-5pt}
\caption{Computational efficiency comparison}
\label{fig:spd}
\end{figure}

We also compare the computational efficiency of our sampling approach
against a naive method that uses all of the original graph to calculate
$\theta$ in an exact fashion.
The results are depicted in Fig.~\ref{fig:spd}.
Obviously, the naive method is very inefficient and our sample-estimate
solution is at least about $50$ times faster with $p=0.3$ on all of the
four graphs.

\header{Validation when graph size is unknown.}
When the graph size is unknown, the second method in
Subsection~\ref{subsec:mle_n_unknown} provides estimates for the number of
nodes with at least one triangle in the graph $\hat{n}_+$ and triadic
cardinality distribution
$\hat\theta^+=(\hat\theta_1^+,\ldots,\hat\theta_W^+)$ for the nodes with at
least one triangle.

\begin{figure}
\vspace{-5pt}
\centering
\subfloat[HepTh]{\includegraphics[width=.5\linewidth]{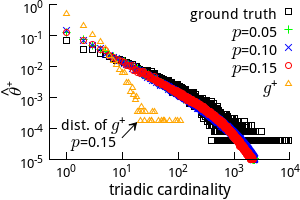}}
\subfloat[DBLP]{\includegraphics[width=.5\linewidth]{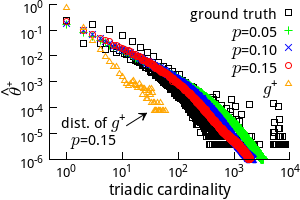}}\\[-8pt]
\subfloat[YouTube]{\includegraphics[width=.5\linewidth]{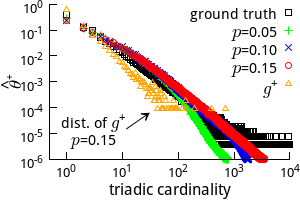}}
\subfloat[Pokec]{\includegraphics[width=.5\linewidth]{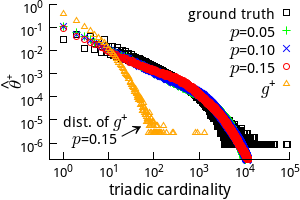}}
\vspace{-5pt}
\caption{Estimates of $\theta^+$ when $|V|$ is unknown. $\hat\alpha$
  for each graph and $p$ ranges from $0.0001$ to $0.01$. $W\!=\!10^4$ and each
  result is averaged over $100$ runs.}
\label{fig:est2}
\vspace{-10pt}
\end{figure}

The results are shown in Fig.~\ref{fig:est2}, using three sample rates
$p=0.05,0.1$ and $0.15$, respectively.
It is clear that the second method also provides good estimates.
Using a fraction of $5\%$ of the data is sufficient to obtain good estimates.
The computational efficiency is similar to results depicted in
Fig.~\ref{fig:spd}.

\begin{figure}
\vspace{-10pt}
\centering
\subfloat[HepTh]{\includegraphics[width=.5\linewidth]{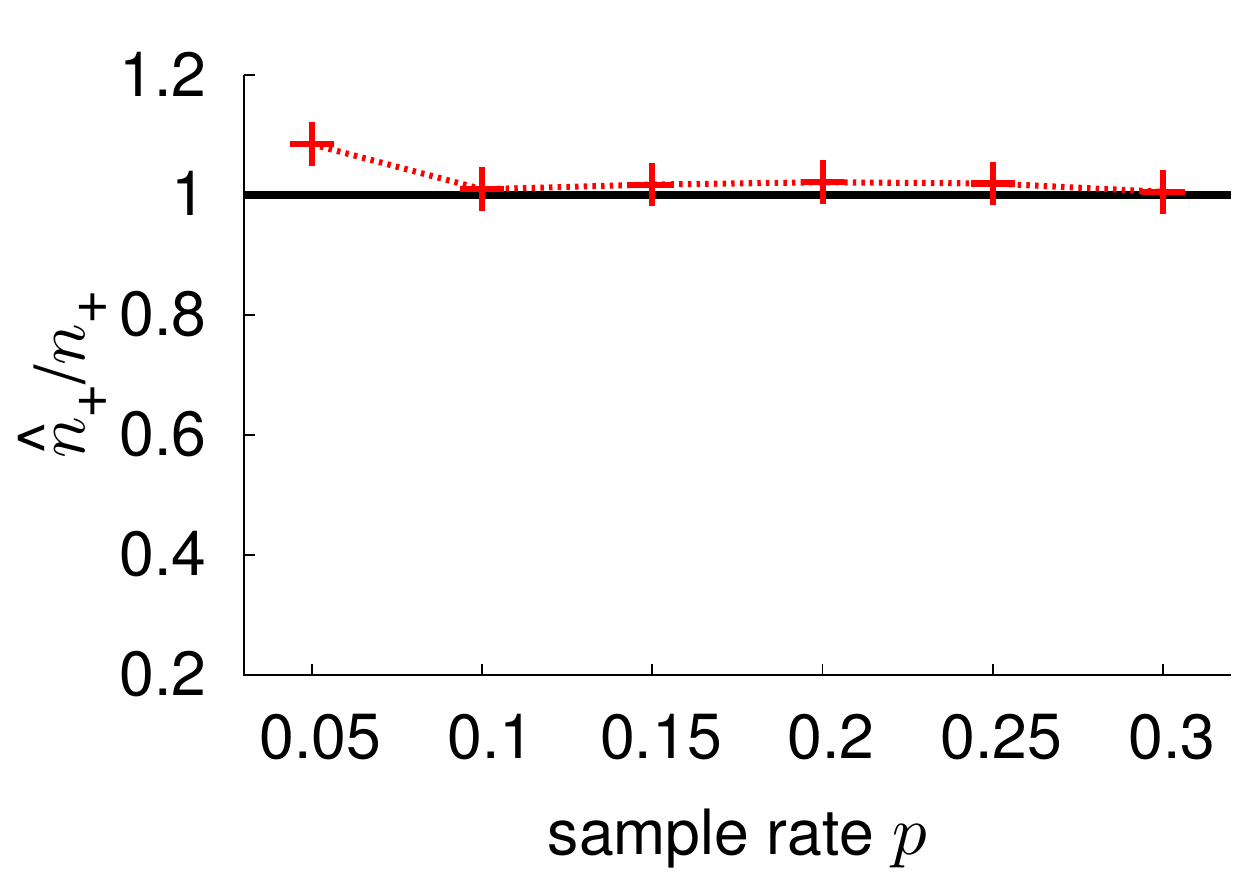}}
\subfloat[DBLP]{\includegraphics[width=.5\linewidth]{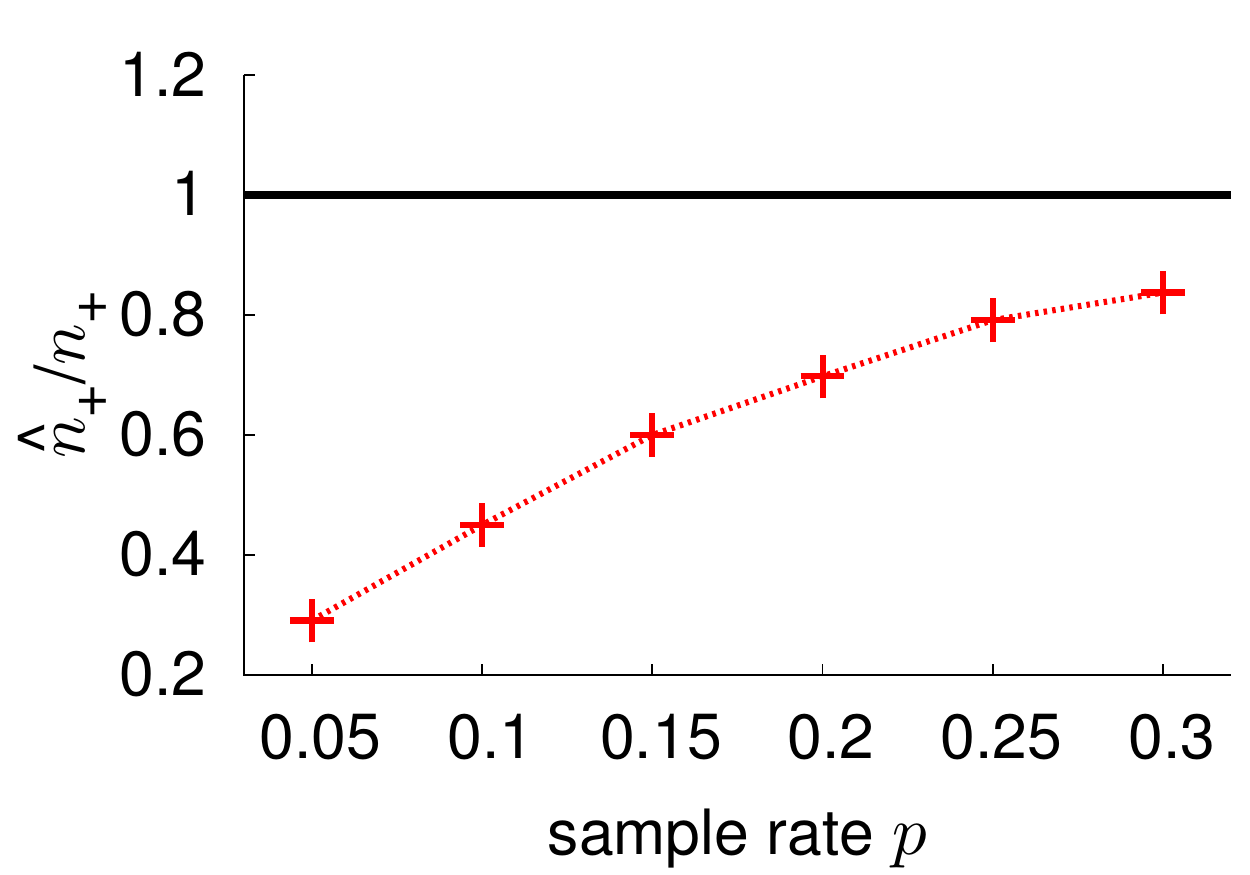}}\\[-10pt]
\subfloat[YouTube]{\includegraphics[width=.5\linewidth]{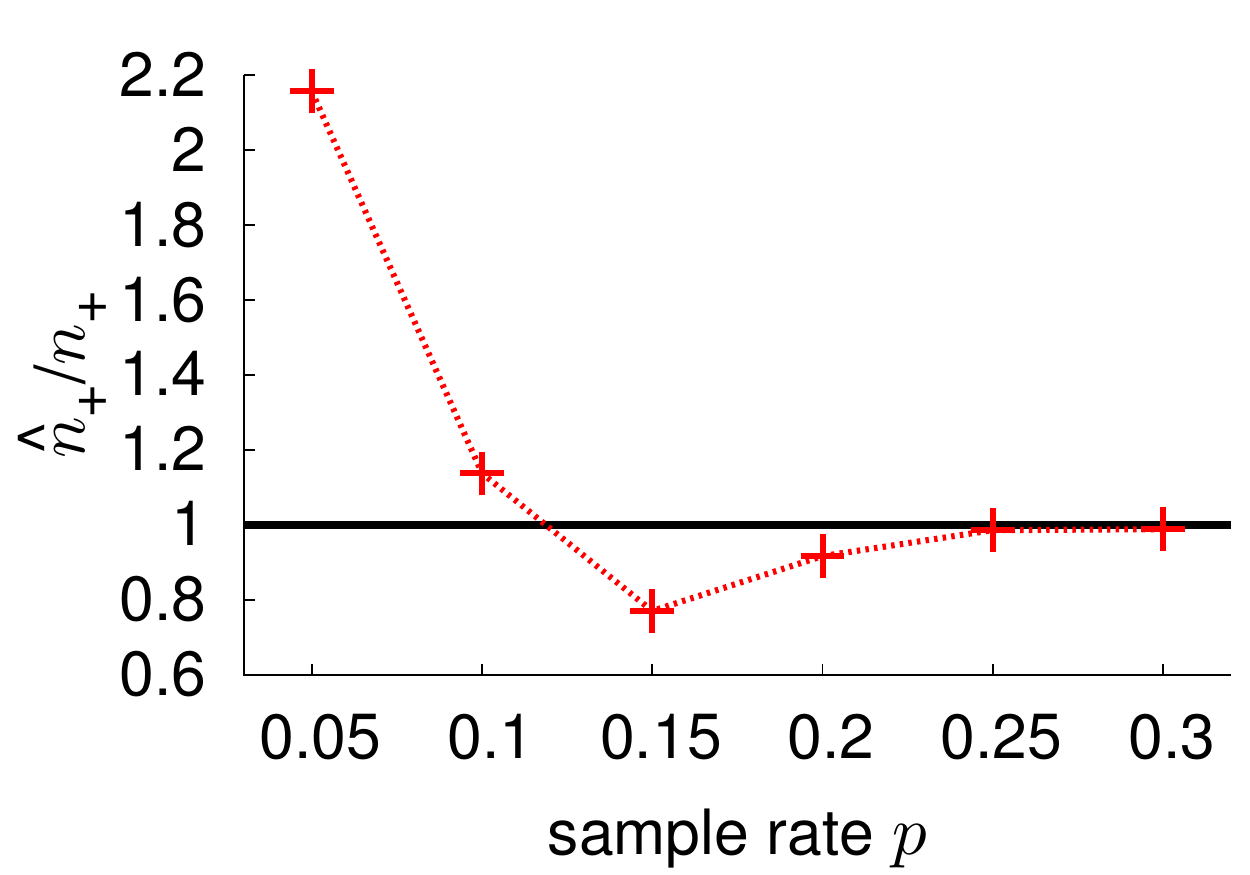}}
\subfloat[Pokec]{\includegraphics[width=.5\linewidth]{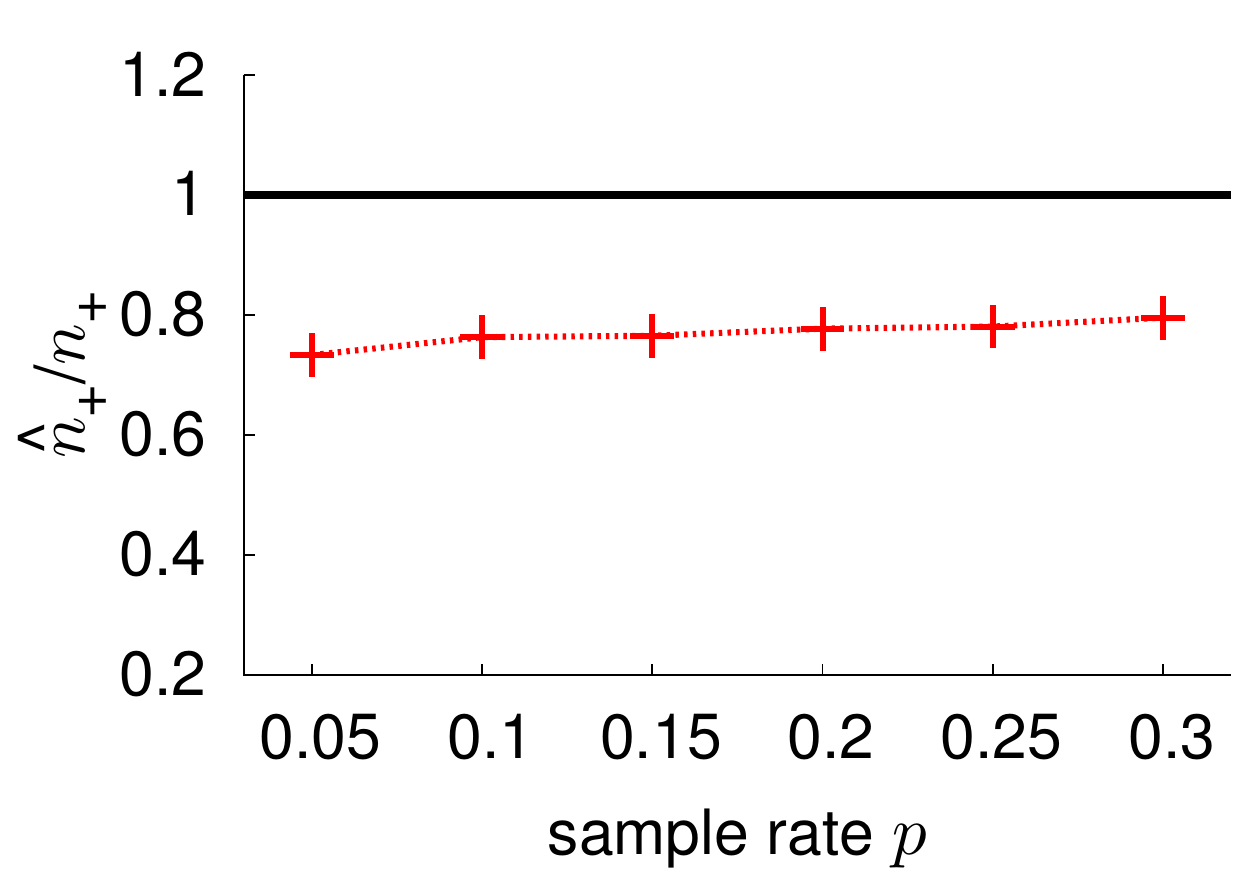}}
\vspace{-5pt}
\caption{Estimates of $n_+$. $W=10^4$ and each result is averaged over
  $100$ runs.}
\label{fig:estn}
\vspace{-10pt}
\end{figure}

The estimate of $n_+$ for each graph is shown in Fig.~\ref{fig:estn}.
Because the majority of the nodes have small triadic cardinalities, good
estimates of $\theta_i^+$ for small values of $i$ are critical for a good
estimate of $n_+$ using estimator~\eqref{eq:nplus}.
For the HepTh graph, estimate $\hat{n}_+$ is very accurate even with small
$p$.
While for the other three graphs, accurate estimates of $n_+$ require
relatively large sample rates, and $\hat{n}_+$ is usually an underestimate
of $n_+$ on DBLP and Pokec due to a slight underestimate of
$\theta_i^+$ for small values of $i$ on the two graphs.
Nevertheless, using a sample rate $p=0.3$, the relative estimation error
for $\hat{n}_+$ is less than $20\%$ for all four graphs.
The design of a better estimator for $n_+$ is left for future work.

\subsection{Application: Burst Detection in 2014 Hong Kong Occupy Central Movement}

In the third experiment, we apply our solution to detect bursts in Twitter
during the 2014 Hong Kong Occupy Central movement.

\header{2014 Hong Kong Occupy Central movement}
a.k.a. the Umbrella Revolution, began in Sept 2014 when activists in Hong
Kong protested against the government and occupied several major streets of
Hong Kong to go against a decision made by China's Standing Committee of
the National People's Congress on the proposed electoral reform.
Protesters began gathering from Sept 28 on and the movement was still
ongoing while we were collecting the data.

\header{Building a Twitter social activity stream.}
The input of our solution is a social activity stream from Twitter.
For Twitter itself, this stream is easily obtained by directly
aggregating tweets of users.
While for third parties who do not own user's tweets, the stream can be
obtained by following users using a set of Twitter accounts, called {\em
detectors}, and aggregating tweets received by detectors (i.e.,
detectors' timelines) to form a social activity stream.
Since the movement had already begun prior to our starting this work, we
rebuilt the social activity stream by searching tweets containing at least
one of the following hashtags: \#OccupyCentral, \#OccupyHK,
\#UmbrellaRevolution, \#UmbrellaMovement and \#UMHK, between Sept 1 and Nov
30 using Twitter search APIs.
This produced $66,589$ Twitter users, and these users form the detectors
from whom we want to detect bursts.
Next, we collect each user's tweets between Sept 1 and Nov 30, and
extract user mentions (i.e., user-user interactions) and user hashtags
(i.e., user-content interactions) from tweets to form a social activity
stream, with a time span of $91$ days.

\header{Settings.}
We set the length of a time window to be one day.
In a time window, we sample each social activity with probability $p=0.3$
and check a social relation with probability $p'=0.3$.
For interaction bursts caused by user-user interactions, because we know
the user population, i.e., $n=66,589$, we apply the first estimation method
to obtain $\hat\theta=(\hat\theta_0,\ldots,\hat\theta_W)$ for each window.
For cascading bursts caused by user-content interactions, as we do not know
the number of hashtags in advance, we apply the second method to obtain
estimates $\hat{n}_+$, i.e., the number of hashtags with at least one
influence triangle, and
$\hat\vartheta^+=(\hat\vartheta_1^+,\ldots,\hat\vartheta_W^+)$ for each
window.
Combining $\hat{n}_+$ with $\hat\vartheta^+$, we use
$\hat{n}_+\hat\vartheta^+$, i.e., frequencies, to characterize patterns of
user-content interactions in each window.
For both $\hat\theta$ and $\hat\vartheta^+$, $W$ is set to be $10^4$.

\begin{figure}
\vspace{-5pt}
\centering
\subfloat[Interaction burst]{
	\includegraphics[width=.5\linewidth]{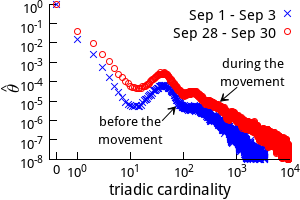}}
\subfloat[Cascading burst]{
	\includegraphics[width=.5\linewidth]{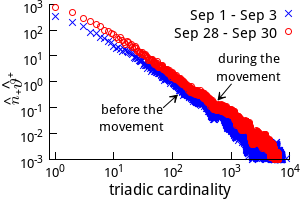}}
\caption{Triadic cardinality distributions before and during the movement.}
\label{fig:diff_week}
\vspace{-10pt}
\end{figure}

\header{Results.}
We first answer the question: are there significant differences for the
two distributions before and during the movement?
In Fig.~\ref{fig:diff_week}, we compare the distributions before (Sept 1
to Sept 3) and during (Sept 28 to Sept 30) the movement.
We can find that when the movement began on Sept 28, the distributions
of the two kinds of interactions shift to the right, indicating that many
interaction and influence triangles form when the movement starts.
Therefore, these observations confirm our motivation for detecting bursts
by tracking triadic cardinality distributions.

\begin{figure*}
\vspace{-10pt}
\centering
\begin{tikzpicture}
\node at (0,0) {\includegraphics[width=.95\linewidth]{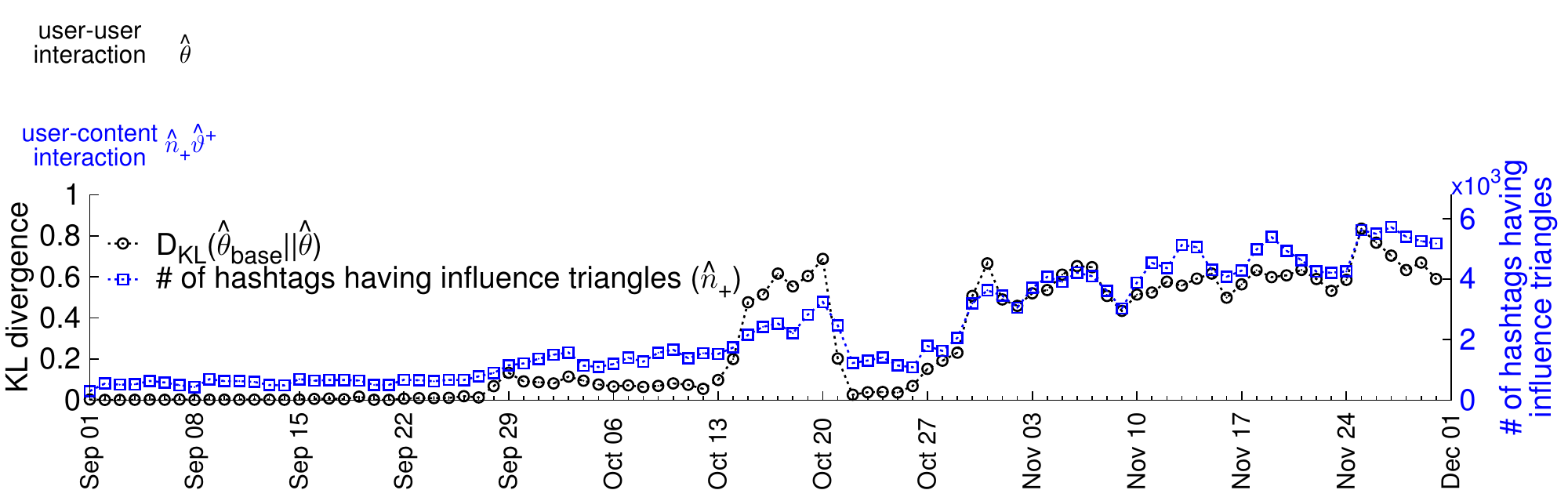}};
\node at (0.4,1.45) {\includegraphics[width=.65\linewidth]{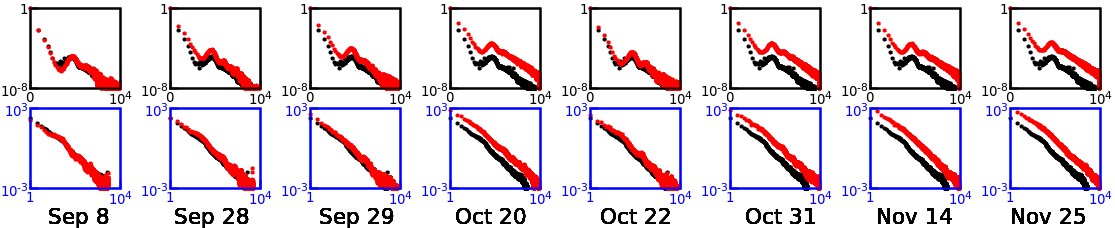}};
\end{tikzpicture}
\vspace{-10pt}
\caption{Burst detection during the 2014 Hong Kong Occupy Central movement
  in Twitter}
\label{fig:kl_hk}
\vspace{-10pt}
\end{figure*}

Next, we track the daily triadic cardinality distributions for the purpose
of burst detection.
To characterize the sudden change in the distributions, we use KL
divergence to calculate the difference between $\hat\theta$
and a base distribution $\theta_\text{base}$.
The base distribution $\theta_\text{base}$ represents a distribution when
the network is dormant, i.e., no bursts are occurring.
For simplicity, we average the triadic cardinality distributions from Sept
1 to Sept 7 to obtain an approximate base distribution
$\hat\theta_\text{base}$, and show the KL divergence
$D_\text{KL}(\hat\theta_\text{base}\parallel\hat\theta)$ in
Fig.~\ref{fig:kl_hk}.

We find that the KL divergence exhibits a sudden increase on Sept 28
when the movement broke out.
The movement keeps going on and reaches a peak on Oct 19 when repeated
clashes happened in Mong Kok at that time.
The movement temporally returned to peace between Oct 22 and Oct 25, and
restarted again after Oct 26.
In Fig.~\ref{fig:kl_hk}, we also show the estimated number of hashtags
having at least one influence triangle.
Its trend is similar to the trend of KL divergence which indicates that the
movement is accompanied with rumors spreading in a word-of-mouth manner.

In conclusion, the application in this section demonstrates that the
using of the triadic cardinality distribution is very useful for
detecting bursts from social activity streams.

\section{Related Work}
\label{sec:related}

Kleinberg first studied this topic in~\cite{Kleinberg2002}, where he used a
multi-state automaton to model a stream consisting of messages.
The occurrence of a burst is modeled by an underlying state transiting into
a bursty state that emits messages at a higher rate than at the non-bursty
state.
Based on this model, many variant models are proposed for detecting bursts
from document streams~\cite{Yi2005,Mathioudakis2010}, e-commerce
queries~\cite{Parikh2008}, time series~\cite{Zhu2003}, and social
networks~\cite{Eftekhar2013}.
Although these models are theoretically interesting, some assumptions made
by them are inappropriate, such as the Poisson process of message arrivals
(see~\cite{Barabasi2005}) and nonexistence of spams/bots, which may limit
their practical usage.

The topic of event detection is also related to our work.
Recently, Chierichetti et al.~\cite{Chierichetti2014} found that Twitter
user tweeting and retweeting count information can be used to detect
sub-events during some large event such as the soccer World Cup of 2010.
Takahashi et al.~\cite{Takahashi2011} proposed a probabilistic model to
detect emerging topics in Twitter by assigning an anomaly score for each user.
Sakaki et al.~\cite{Sakaki2010} proposed a spatiotemporal model to detect
earthquakes using tweets.
Different from theirs, we exploit the triangle structure existing in user
interactions which is robust against common spams and can be efficiently
estimated using our method.

The triangle structure can be considered as a type of network motif, which
is introduced in~\cite{Milo2002} when the authors were studying how to
characterize structures of different types of networks.
Turkett et al.~\cite{Turkett2011} used motifs to analyze computer network
usage, and \cite{Wang2014} proposed sampling methods to efficiently
estimate motif statistics in a large graph.
However, both the motivation in \cite{Turkett2011} and subgraph statistics
defined in \cite{Wang2014} are different from ours.

Recently, there are many works on estimating the number of
triangles~\cite{Tsourakakis2009,Budak2011,Pavan2013,Jha2013,Ahmed2014} or
clustering coefficient~\cite{Seshadhri2013} in a large graph.
However, these methods cannot be used to estimate the triadic cardinality
distribution.
Becchetti et al.~\cite{Becchetti2008} used a min-wise hashing method to
approximately count triangles for each individual node in an undirected
simple graph.
Our method does not rely on counting triangles for each individual node.
Rather, we use a carefully designed estimator to estimate the statistics
from a sampled graph, which is demonstrated to be efficient and accurate.

\section{Conclusion}
\label{sec:conclusion}

Online social networks provide various ways for users to interact with
other users or media content over the Internet, which bridge
the online and offline worlds tightly.
This provides an opportunity to researchers to leverage online user
interactions so as to detect bursts that may cause impact to the offline
world.
We find that the emergence of bursts caused by either user-user
interaction or user-content interaction are accompanied with the formation
of triangles in users' interaction networks.
This finding prompts us to devise a new method for burst detection in OSNs by
introducing the triadic cardinality distribution.
Triadic cardinality distribution is demonstrated to be robust against common
spams which makes it a more suitable indicator for detecting bursts than the
volume of user activities.
We design a sample-estimate solution that can efficiently and accurately
estimate triadic cardinality distribution from high-speed social activity
streams in a near-real-time fashion, which makes it applicable in practice.

\bibliographystyle{abbrv}

\end{document}